\documentclass[twocolumn,aps,superscriptaddress]{revtex4}
\usepackage{graphicx}
\usepackage{float}
\usepackage{dcolumn}
\usepackage{bm}
\usepackage{color}
\usepackage{amsmath}
\usepackage{amssymb}
\usepackage{amsfonts}
\usepackage{esint}
\usepackage{times}
\usepackage{xcolor}
\usepackage{phaistos}
\usepackage{braket}
\usepackage{comment}
\usepackage{multirow}

\usepackage{pdfpages}

\usepackage[colorlinks,linkcolor=blue,anchorcolor=blue,urlcolor=blue,urlcolor=blue,citecolor=blue]{hyperref}

\begin{document}

\title{Scalable fluxonium qubit architecture with tunable interactions \\ between non-computational levels}

\author{Peng Zhao}
\email{shangniguo@sina.com}
\affiliation{Hefei National Laboratory, Hefei 230088, China}
\author{Guming Zhao}
\author{Shaowei Li}
\author{Chen Zha}
\author{Ming Gong}
\email{minggong@ustc.edu.cn}
\affiliation{Hefei National Research Center for Physical Sciences at the Microscale and School of Physical Sciences, University of Science and Technology of China, Hefei 230026, China}
\affiliation{Shanghai Research Center for Quantum Science and CAS Center for Excellence in Quantum Information and Quantum Physics, University of Science and Technology of China, Shanghai 201315, China}
\affiliation{Hefei National Laboratory, Hefei 230088, China}

\date{\today}

\begin{abstract}

The fluxonium qubit has emerged as a promising candidate for superconducting quantum computing due to
its long coherence times and high-fidelity gates. Nonetheless, further scaling up and improving performance
remain critical challenges for establishing fluxoniums as a viable alternative to transmons. A key
obstacle lies in developing scalable coupling architectures. In this work, we introduce a scalable fluxonium
architecture that enables decoupling of qubit states while maintaining tunable couplings
between non-computational states. Beyond the well-studied ZZ crosstalk, we identify that always-on
interactions involving non-computational levels can significantly degrade the fidelities of initialization,
control, and readout in large systems, thereby impeding scalability. Based on two possible
physical realizations of the architecture, we demonstrate that the issue can
be mitigated by implementing tunable couplings for fluxonium plasmon transitions, meanwhile enabling
fast, high-fidelity gates with passive ZZ suppression. This comparative analysis enables us to establish
general principles for realizing the architecture while understanding and addressing implementation-specific challenges.
\end{abstract}

\maketitle


\section{Introduction}\label{SecI}

Superconducting qubits~\cite{Kjaergaard2020,Gyenis2021}, particularly transmons~\cite{Koch2007}, are a
leading platform~\cite{Krantz2019} for scalable quantum processors capable of tackling classically
intractable problems. Over the past two decades, transmons have enabled most advances in superconducting quantum computing~\cite{Wendin2017,Huang2020,Blais2021,Kwon2021,Cheng2023}.
Meanwhile, fluxonium qubits~\cite{Manucharyan2009,note2025} have emerged as a promising alternative due to superior coherence
times~\cite{Nguyen2019} (reaching milliseconds~\cite{Somoroff2023,Wang2024}) and strongly anharmonic spectrum
with rich level structures~\cite{Manucharyan2012}, enabling fast, low-leakage gates~\cite{Nesterov2018,Ficheux2021,Xiong2022,Nesterov2021,Nesterov2022,Dogan2023,Lin2024,Bao2022,Chen2022,Ma2024,Simakov2023,
Ding2023,Rosenfeld2024,Moskalenko2021,Moskalenko2022,Weiss2022,Zhang2024,Rower2024}. Recent experiments
have demonstrated single- and two-qubit gate errors below $10^{-4}$~\cite{Somoroff2023,Ding2023,Rower2024}
and $10^{-3}$~\cite{Lin2024,Ding2023,Zhang2024}, respectively, highlighting fluxonium's potential for
fault-tolerant superconducting quantum computing.

For fluxoniums to become a viable alternative to transmons, further scaling and improving performance are
essential. As the transmon's development highlights, a critical challenge lies in realizing scalable coupling
architectures, which can address issues including quantum crosstalk~\cite{Blais2021,DiCarlo2009,Mundada2019}, frequency crowding~\cite{Brink2018,Kelly2015,Chen2014}, and control crosstalk~\cite{Chen2014,Orlando1999}. While
several fluxonium coupling schemes have been demonstrated, their scalability remains uncertain. Direct capacitive~\cite{Nesterov2018,Ficheux2021,Xiong2022,Nesterov2021,Nesterov2022,Dogan2023,Bao2022,Chen2022}
or inductive~\cite{Nesterov2018,Ma2024} coupling suffers from residual ZZ interactions and frequency crowding that
hinder scaling-up. And approaches using multi-path couplings~\cite{Nguyen2022,Lin2024} or
inductive couplers~\cite{Weiss2022,Zhang2024} with shared inductances, while potentially addressing
issues involving unwanted interactions, introduce intrinsic flux crosstalk~\cite{Orlando1999,Paauw2009}
that compromises precise control for high-fidelity gates at scale.

Consequently, indirect capacitive couplings might be more scalable approaches. While tunable coupling
via an ancilla fluxonium has been demonstrated~\cite{Moskalenko2021,Moskalenko2022} (analogous to
transmon's~\cite{Yan2018}), the fluxonium's weak transition dipole ($\sim10\times$ smaller than transmons)
in qubit subspace leads to inefficient coupling, thus compromising gate speed and ZZ suppression. Alternative
architectures employing transmon-based couplers can suppress ZZ crosstalk and enable coupler-mediated
interactions between fluxonium's plasmon transitions (leveraging their larger dipoles comparable
to transmons) for microwave-activated CZ gates~\cite{Ding2023}.

\begin{figure}[tbp]
\begin{center}
\includegraphics[keepaspectratio=true,width=\columnwidth]{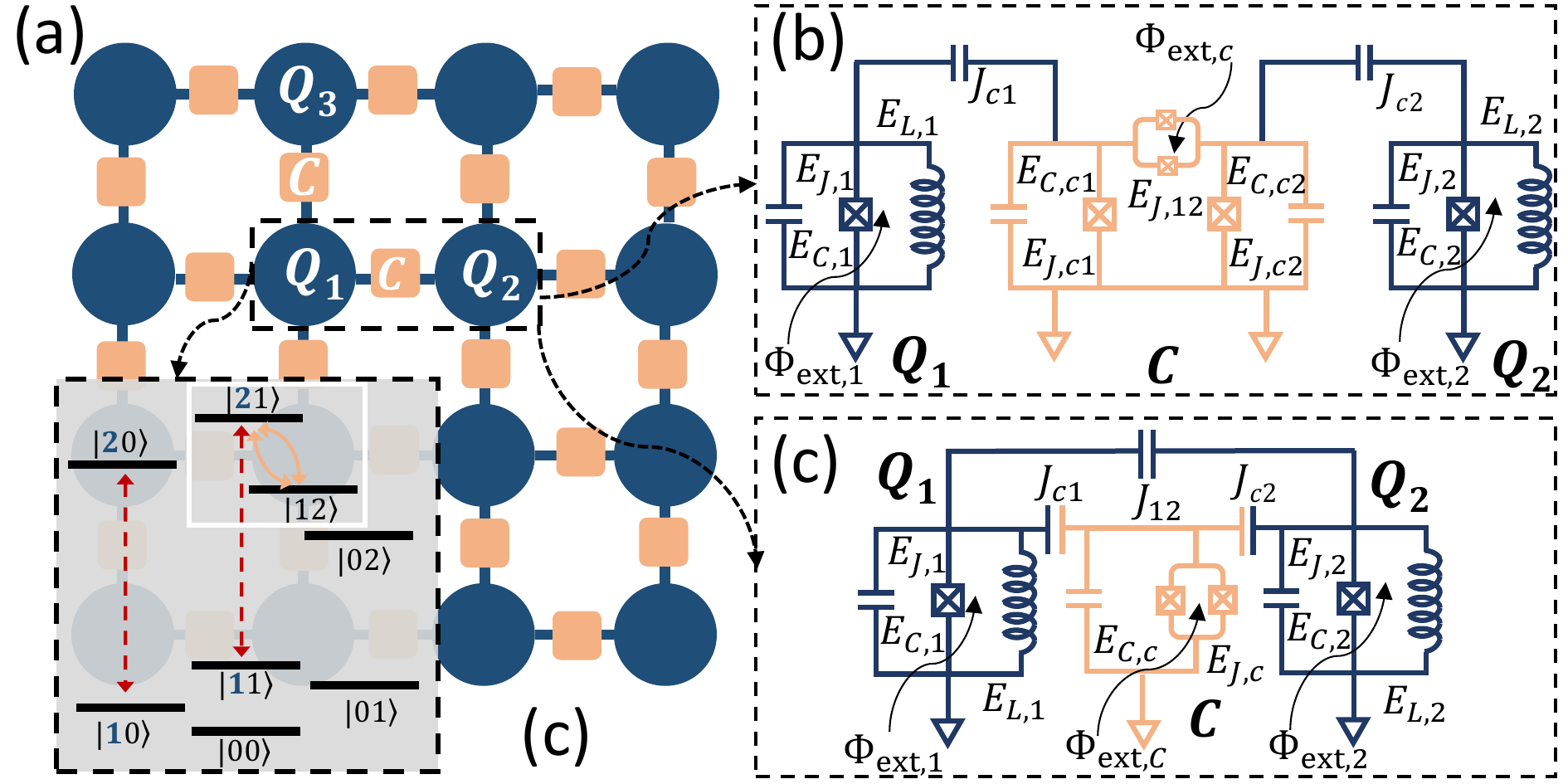}
\end{center}
\caption{(a) A qubit lattice comprising fluxoniums (circles) coupled via couplers (squares).
The inset depicts the energy levels for coupled fluxoniums ($Q_{1}$ and $Q_{2}$), showing that flip-flop
interactions (curved arrows, $|12\rangle\leftrightarrow|21\rangle$) between plasmon
transitions ($|1\rangle\leftrightarrow|2\rangle$) cause state-dependent plasmon frequency shifts ($Q_{1}$)
through the interaction-induced level repulsion. This effect enables selective driving of
transitions (e.g., $|11\rangle\leftrightarrow|21\rangle$) for CZ gates between $Q_{1}$ and $Q_{2}$,
yet it also introduces quantum crosstalk, changing the gate frequency depending on the states of
neighbor-coupled spectators, e.g., $Q_{3}$. (b) A fluxonium architecture featuring tunable
plasmon interactions, where fluxoniums are coupled capacitively to a double-transmon
coupler (comprising transmons coupled via a tunable inductor, i.e., a dc SQUID) or a
single-transmon coupler (c).}
\label{fig1}
\end{figure}

However, two fundamental limitations emerge: (1) State-dependent plasmon
frequency shifts due to the level repulsion from the plasmon interactions, i.e., plasmon frequencies of each fluxonium
depend on its coupled neighbors's states. As illustrated in Fig.~\ref{fig1}(a), this shift presents both opportunities
and challenges for qubit control. It enables selective driving of gate transitions (e.g., $|11\rangle\rightarrow|21\rangle$,
hereafter, the system state is denoted as $|Q_{1}Q_{2}\rangle$)
for realizing CZ gates (e.g., between $Q_{1}$ and $Q_{2}$)~\cite{Nesterov2018,Ficheux2021,Ding2023}, yet it also introduces
quantum crosstalk, which changes the gate transition frequency of interest depending on the states of neighbor-coupled
spectators (e.g., $Q_{3}$). (2) Frequency collisions involving near-resonant plasmons and couplers for gates temporarily occupying
non-computational states (e.g., high-energy levels~\cite{Nesterov2018,Ficheux2021,Ding2023} or coupler
levels~\cite{Simakov2023,Ding2023,Rosenfeld2024}), giving rise to the frequency crowding problem. These effects critically
constrain gate fidelities in multiqubit systems. Similarly, the fidelities of qubit readout~\cite{Nguyen2022,Stefanski2024} and
initialization~\cite{Manucharyan2009b,Zhang2021,Wang2024a} are degraded when involving plasmon
transitions. These intrinsic limitations render the architecture with always-on plasmon interactions impractical
for large-scale fluxonium processors.

To this end, here we exploit fluxonium's rich level structures and strong anharmonicity to develop a scalable architecture with tunable
plasmon interactions via couplers. Crucially, the small transition dipoles within computational
subspace (qubit transitions) strongly suppress coupler-mediated qubit interactions. Meanwhile,
the coupler can mediate tunable flip-flop interactions between plasmon transitions (facilitated by
their transmon-like dipoles), allowing fast CZ gates and mitigation of quantum crosstalk and frequency
crowding from fixed couplings to spectators. The architecture can serve as a foundation
for larger-scale fluxonium processors, while the underlying design principle offer general guidelines
for future scalable designs.

\section{circuit model and system Hamiltonian}\label{SecII}

To start, we note that fluxonium plasmon transitions exhibit transmon-like dipoles, thus
tunable couplers originally developed for transmons~\cite{Mundada2019,Yan2018,Goto2022,Campbell2023,Egorova2024,Wang2024} could in principle be
used for realizing tunable plasmon interactions. However, since fluxoniums exhibit complex
level structures with transitions spanning $\sim100$ MHz to
over 10 GHz, the key challenge for employing these designs lies in achieving tunable
interactions while ensuring compatibility with fast, high-fidelity control of high-coherence
fluxoniums. For illustration purposes, this work focuses on two specific schemes, wherein
fluxoniums interact via a two-mode coupler~\cite{Miyanaga2021,Goto2022,Campbell2023} or a single-mode
coupler~\cite{Yan2018}. This comparative study aims to elucidate general principles for implementing the
proposed architecture while identifying and resolving implementation-specific challenges.

Figures~\ref{fig1}(b) and~\ref{fig1}(c) present the two implementations of the architecture,
where fluxoniums are coupled via a double-transmon coupler (DTC) and a single-transmon
coupler (STC), respectively. For the DTC, which comprises two transmons coupled inductively~\cite{Geller2015,Dumur2015,Chen2014,Kounalakis2018}
via a dc SQUID (a tunable inductor), the flux bias modulates
the intermode inductive coupling, see Appendix~\ref{IA}, while for the STC, the flux bias tunes
the coupler frequency itself. The two setups are modeled by the following Hamiltonians (hereafter, $\hbar=1$)
\begin{equation}
\begin{aligned}\label{eq1}
&H^{(D)}=H^{(F)}+\sum_{i=1,2} [4 E_{C,ci} \hat n^2_{ci} - E_{J,ci}\cos\hat\varphi_{ci}+J_{ci}\hat n_i \hat n_{ci}]\\
&\quad\quad\quad-E_{J,12}\cos(\frac{\varphi_{\text{ext},c}}{2})\cos(\hat\varphi_{c1}-\hat\varphi_{c2}),\\
&H^{(S)}=H^{(F)}+4 E_{C,c} \hat n^2_{c} - E_{J,c}\cos(\frac{\varphi_{\text{ext},c}}{2})\cos\hat\varphi_{c}\\
&\quad\quad\quad+J_{c1}\hat n_1 \hat n_{c}+J_{c2}\hat n_2 \hat n_{c}+J_{12}\hat n_1 \hat n_{2}
\end{aligned}
\end{equation}
with the fluxonium Hamiltonian $H^{(F)}=\sum_{i=1,2}[4 E_{C,i} \hat n^2_i + E_{L,i}(\hat\varphi_i - \varphi_{\text{ext},i})^2/2 - E_{J,i}\cos\hat\varphi_i]$ in the irrotational gauge~\cite{You2019,Riwar2022,Osborne2024,Bryon2023}. Here, $E_C$, $E_J$, and $E_L$ are the charging, Josephson, and inductive energies, respectively,
subscripts $i$ and $ci\,(c)$ label the fluxonium and the transmon of the DTC (STC), and $\varphi_\text{ext}$
is the external phase bias with $\varphi_\text{ext}=2\pi\Phi_\text{ext}/\Phi_0$. For clarity, we consider that
the SQUID's two junctions are identical for both setups and flux crosstalk between the main loop and the SQUID loop in
the DTC are compensated for allowing independent flux control, see Appendix~\ref{IA}.

After approximating transmons as harmonic modes, i.e., introducing
$\hat \varphi_{ci} = \varphi_{ci,{\rm zpf}}(\hat a_{ci}^{\dag}+\hat a_{ci})$ and
$\hat n_{ci} = i n_{ci,{\rm zpf}}(\hat a_{ci}^{\dag}-\hat a_{ci})$ with the phase (number) zero-point fluctuation $\varphi_{ci,{\rm zpf}}=(8E_{C,ci}/E_{J,ci})^{1/4}/\sqrt{2}$ ($n_{ci,{\rm zpf}}=1/[2\varphi_{ci,{\rm zpf}}]$)~\cite{Koch2007}, and focusing on
one specific fluxonium's transition $|k\rangle\leftrightarrow|l\rangle$, the Hamiltonians
can be approximated by (see Appendixes~\ref{IB} and~\ref{IIA})
\begin{equation}
\begin{aligned}\label{eq2}
&H_{kl}^{(D)}=\sum_{i=1,2}[\omega_{kl,i}\hat\sigma_{kl,i}^{\dag}\hat\sigma_{kl,i}+g_{kl,i}(\hat\sigma_{kl,i}^{\dag}\hat a_{ci}+h.c.)
\\&\quad\quad\quad+\omega_{ci}\hat a_{ci}^{\dag}\hat a_{ci}]+g_{c}(\hat a_{c1}+\hat a_{c1}^{\dag})(\hat a_{c2}+\hat a_{c2}^{\dag}),\\
&H_{kl}^{(S)}=\sum_{i=1,2}[\omega_{kl,i}\hat\sigma_{kl,i}^{\dag}\hat\sigma_{kl,i}+g_{kl,i}(\hat\sigma_{kl,i}^{\dag}\hat a_{c}+h.c.)]
\\&\quad\quad\quad+\omega_{c}\hat a_{c}^{\dag}\hat a_{c}+g_{kl,12}(\hat\sigma_{kl,1}^{\dag}\hat\sigma_{kl,2}+h.c.)
\end{aligned}
\end{equation}
where $\hat\sigma_{kl,i}=|k\rangle\langle l|_{i}$ ($\sigma_{kl,i}^{\dag}$) is the lowering (raising)
operator for the transition of $\omega_{kl,i}$, $a_{ci(c)}$ ($a_{ci(c)}^{\dag}$) denotes the destroy (creation)
operator for the coupler mode of $\omega_{ci(c)}$, and $g_{kl,i}=J_{ci} \langle k1|\hat n_i \hat n_{ci(c)}|l0\rangle$
($g_{kl,12}=J_{12} \langle kl|\hat n_1 \hat n_{2}|lk\rangle$) represents the fluxonium-coupler (the fluxonium-fluxonium) interaction
strength. Here, $g_{c}=-E_{J,12}\cos(\varphi_{\text{ext},c}/2)\varphi_{c1,{\rm zpf}}\varphi_{c2,{\rm zpf}}$ is the DTC's intermode inductive
coupling strength, see Appendix~\ref{IB}.

Assuming degenerate modes ($\omega_{ci}=\omega_{c}$) for the DTC and the interaction
strength $g_{kl,i}$ significantly smaller than the fluxonium-coupler detuning $\Delta_{kl,i}=\omega_{kl,i}-\omega_{c}$
for both setups, an effective Hamiltonian can be derived by eliminating the direct fluxonium-coupler
interactions, yielding (see Appendixes~\ref{IB} and~\ref{IIA}):
\begin{equation}
\begin{aligned}\label{eq3}
&H_{kl}^{({\rm eff})}= \sum_{i=1,2}[\omega_{kl,i}\hat\sigma_{kl,i}^{\dag}\hat\sigma_{kl,i}]+g_{kl,{\rm eff}}(\hat\sigma_{kl,1}^{\dag}\hat\sigma_{kl,2}+h.c.),
\end{aligned}
\end{equation}
where $g_{kl,{\rm eff}}$ is the coupler-mediated flip-flop interaction strength with
\begin{equation}
\begin{aligned}\label{eq4}
&g_{kl,{\rm eff}}^{(D)}\approx\frac{g_{kl,1}g_{kl,2}g_{c}}{2}\sum_{i=1,2}\left(\frac{1}{\Delta_{kl,i}^{2}}
+\frac{\Delta_{kl,i}/\omega_{c}}{\Delta_{kl,i}^{2}}\right),
\\&g_{kl,{\rm eff}}^{(S)}\approx g_{kl,12}+\frac{g_{kl,1}g_{kl,2}}{2}\left(\frac{1}{\Delta_{kl,1}}+\frac{1}{\Delta_{kl,2}}\right),
\end{aligned}
\end{equation}
for the DTC and STC setup, respectively.

Equation~(\ref{eq4}) indicates that the small qubit transition dipole effectively
suppresses coupler-mediated qubit interactions, while the plasmon's transmon-like dipole enables
strong tunability of plasmon interactions. This allows programmable control of the plasmon
interactions through either flux control of the DTC's intermode coupling or by tuning
the STC frequency, without compromising the decoupling of qubit states. These findings establish general
principles for implementing the proposed architecture, by leveraging the contrasting dipole between
qubit and plasmon transitions, any capacitive-based tunable
coupler~\cite{Mundada2019,Yan2018,Miyanaga2021,Goto2022,Campbell2023,Egorova2024,Wang2024}) with modes matching
fluxonium plasmons could enable tunable plasmon interactions while
suppressing qubit interactions.

Moreover, Equation~(\ref{eq4}) also reveals a fundamental distinction between the DTC and STC setup. In the DTC setup, the
interaction nulling condition depends solely on the intermode coupling and is irrelevant to fluxoniums (transition
frequencies and dipoles). This enables simultaneous nulling of all plasmon interactions at a single bias point, contrasting
sharply with the STC setup where the nulling typically requires different biases for different transitions. However, this ideal
behavior strictly holds only in the absence of intermode capacitive coupling. In practical devices, parasitic capacitive
couplings inevitably exist, with the dominant contribution arising from the self-capacitance of the coupling junction. This
introduces an additional coupling term $J_{C,12} \hat n_{c1} \hat n_{c2}$ that renders the nulling condition
frequency-dependent, typically requiring distinct biases to null different interactions, see Appendix~\ref{IC1}.

\begin{table}[!htb]
\caption{\label{tab:parameters} Hamiltonian parameters of the coupled
circuit depicted in Fig.~\ref{fig1}. All parameters are the same for both
setups except for the parenthetical values used for the STC setup.}
\begin{ruledtabular}
\begin{tabular}{cccc}
$ $&
$E_C$ (GHz)&
$E_L$ (GHz)&
$E_J$ (GHz)\\\hline
Fluxonium $Q_{1}$ & 1.41 & 0.80 & 6.27  \\
Fluxonium $Q_{2}$ & 1.30 & 0.59 & 5.71 \\
Spectator $Q_{3}$ & 1.33 & 0.60 & 5.40 \\
Transmon  & 0.25 (0.32) & $-$ & 9 (55) \\
\hline
\hline
$ $ & $J_{c1 (c2)}$ (MHz) & $J_{C,12 (12)}$ (MHz)  & $E_{J,12}$ (GHz)\\\hline
Coupling strengths & 550 (500) & 100 (125)  & 7
\end{tabular}
\end{ruledtabular}
\end{table}

\section{Tunable plasmon interactions with feasible circuit parameters}\label{SecIII}

\begin{figure}[tbp]
\begin{center}
\includegraphics[keepaspectratio=true,width=\columnwidth]{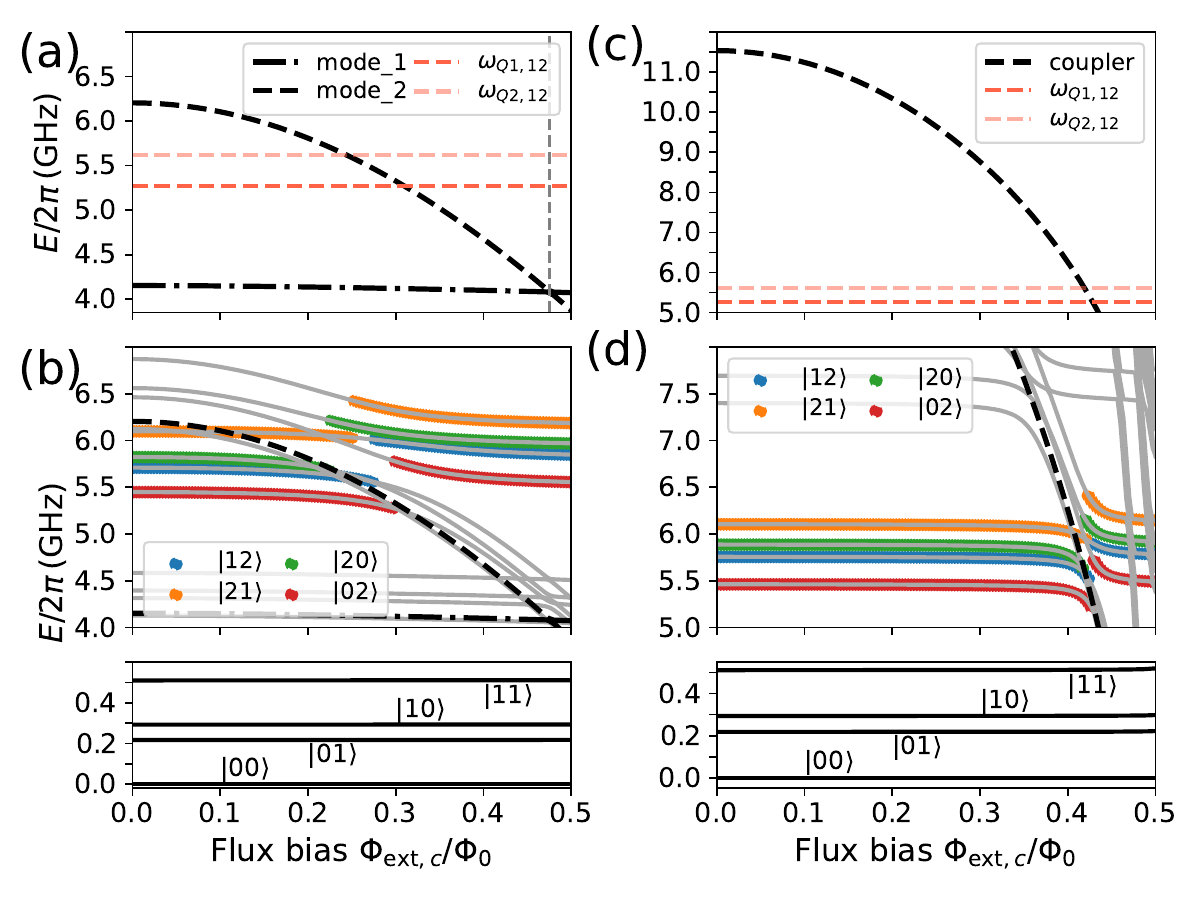}
\end{center}
\caption{Coupler's eigenmode frequencies and energy levels of the full coupled circuit versus
the coupler flux bias. (a,b) and (c,d) are for the DTC and STC setups, respectively.
In (a) and (c), horizontal dashed lines denote the fluxonium's bare plasmon transition ($|1\rangle\leftrightarrow|2\rangle$)
frequencies and vertical dashed lines indicate the bias for nulling the
DTC's intermode coupling (excluding the fluxoniums).}
\label{fig2}
\end{figure}

Following the analytical treatment, we now evaluate their realization
with the experimentally realistic parameters specified in Table~\ref{tab:parameters} (primarily
adopted from Ref.~\cite{Ding2023}). To examine their compatibility with
fast, high-fidelity control of high-coherence fluxoniums, here we mainly focus on
the qubit transition and the plasmon transition $|1\rangle\leftrightarrow|2\rangle$, see
Fig.~\ref{fig1}(a).

For the DTC setup, Figure~\ref{fig2}(a) shows the flux dependence of coupler's eigenmode
frequencies, with horizontal dashed lines indicating the bare
plasmon frequencies. When including intermode capacitive coupling, the intermode coupling nulling point shifts slightly from
the predicted bias at $\varphi_{\text{ext},c}=\pi$, see Eq.~(\ref{eq1}) and the bias indicated by the vertical
dashed line. Figure~\ref{fig2}(b) presents the system spectrum, uncovering
two key features: (1) The computational subspace exhibits negligible interaction with the coupler, evidenced
by the almost flux-independent qubit spectrum. This mainly results from the weak qubit's transition dipole,
which minimizes coupler-induced decoherence (see Appendixes~\ref{IF1} and~\ref{IIB}) and preserves the high-coherence
in qubit subspace. (2) Due to the plasmon's large dipole, there exist strong level repulsions from
the plasmon-coupler interactions, revealing its capability to mediate strong plasmon interactions. Similar
behavior is also obtained for the STC setup, as shown in Figs.~\ref{fig2}(c) and~\ref{fig2}(d).

\begin{figure}[tbp]
\begin{center}
\includegraphics[keepaspectratio=true,width=\columnwidth]{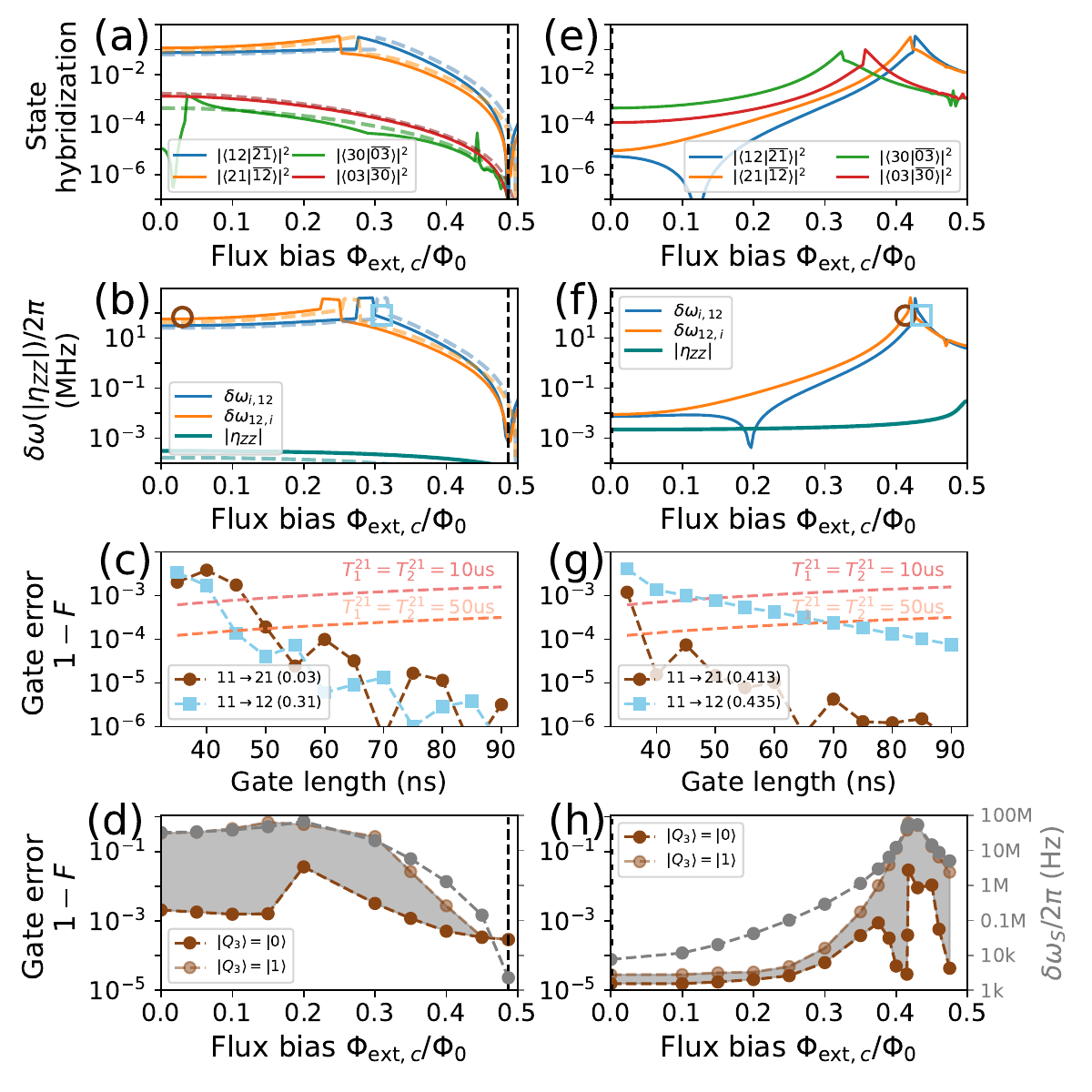}
\end{center}
\caption{Coupler-mediated plasmon interactions, quantified by state hybridization (a,e) and state-dependent plasmon
frequency shifts (b,f), versus the coupler flux bias (sudden jumps in curves
are caused by state labeling failure near avoided crossings), CZ gate errors with varying gate lengths (c,g), and spectator-induced gate
errors with varying residuals (d,h). (a-d) and (e-h) are for the DTC and STC setup, respectively. In (b,f), vertical dashed lines
mark coupler idle points (minimizing the shifts) while open circles/squares indicate coupler interaction points for
CZ gates. In (c,g), the gate lengths exclude the 3 ns ramp time for biasing the coupler from the
idle point to the interaction point, and dashed lines are calculated gate errors from the relaxation
and dephasing of noncomputational gate levels. (d,h) grey dots show the shift in gate frequency due to the
coupling between $Q_{1}$ and the spectator $Q_{3}$, see Fig.~\ref{fig1}(a), and brown dots show the 50-ns gate error with $Q_{3}$
prepared in different states.}
\label{fig3}
\end{figure}

Since fluxoniums are commonly biased at their half-flux sweet spots, here we employ two complementary metrics to quantify the
coupler-mediated plasmon interactions instead of extracting its strengths directly from energy-level splitting: (1) Interaction-induced
state hybridization between system eigenstates and bare states~\cite{Heunisch2023}, quantified
by $|\langle kl|\overline{mn}\rangle|^{2}$, where $|kl\rangle$ denotes
the system eigenstate adiabatically connected to the bare state $|\overline{kl}\rangle$~\cite{Nesterov2018,Galiautdinov2012}.
(2) State-dependent plasmon frequency shifts arising from interaction-induced level repulsions, see Fig.~\ref{fig1}(a), i.e.,
\begin{equation}
\begin{aligned}\label{eq5}
&\delta\omega_{12,i}=|(E_{21}-E_{11})-(E_{20}-E_{10})|,\\
&\delta\omega_{i,12}=|(E_{12}-E_{11})-(E_{02}-E_{01})|,
\end{aligned}
\end{equation}
for the two fluxoniums, respectively, where $E_{kl}$ denote the energy of the
eigenstate $|kl\rangle$. The minimization of both quantities indicates decoupling of plasmon
transitions and suppression of associated interactions.

For the DTC setup, Figures~\ref{fig3}(a) and~\ref{fig3}(b) show the flux dependence of state
hybridization and frequency shift for the plasmon transition $|1\rangle\leftrightarrow|2\rangle$, comparing cases
with (solid) and without (dashed) intermode capacitive coupling. Both quantities show consistent flux dependence
and simultaneously minimize at a predicted flux bias (Fig.~\ref{fig2}(a)), confirming the programmable
control and suppression of plasmon interactions. The ZZ coupling strength $\eta_{ZZ}=(E_{11}-E_{01})-(E_{10}-E_{00})$
remains below 1 kHz across the entire bias range, verifying effective decoupling of qubit states. While intermode capacitive
coupling renders the nulling condition plasmon-dependent, see the state hybridization associated
with $|0\rangle\leftrightarrow|3\rangle$ in Fig.~\ref{fig3}(a), the observed dependence is negligible for practical
applications. However, this behavior shows significant dependence on coupler frequencies, see Appendixes~\ref{IE}.

As discussed before, the STC setup requires distinct biases to null different transitions, see
Figs.~\ref{fig3}(e) and~\ref{fig3}(f). However, when biasing the coupler near its maximum
frequency, all interactions can be simultaneously suppressed to practically acceptable levels (e.g., $\delta\omega<10\,\rm kHz$).
This offers significant advantages over the DTC setup by eliminating the need for
precise coupler tuning to suppress plasmon interactions, and thus providing inherent tolerance against frequency
collisions and fabrication-induced parameter variations.

Leveraging tunable plasmon interactions, one can engineer state-dependent frequency shifts to realize
microwave-activated CZ gates~\cite{Nesterov2018,Ficheux2021}. As shown in Fig.~\ref{fig1}(a), the plasmon interaction
of $|1\rangle\leftrightarrow|2\rangle$ can enable selective excitation
of $|11\rangle\leftrightarrow|21(12)\rangle$ and $|10(01)\rangle\leftrightarrow|20(02)\rangle$. Here, CZ gates can be realized by
firstly tuning the coupler from its idle point (vertical dashed lines) to the interaction point (e.g., open circles/squares), then
waiting for one period of the selectively driving Rabi oscillation, and finally biasing the coupler back
to the idle point (see Appendix~\ref{IIIA}). By avoiding collisions with other parasitic transitions (e.g., $|21(12)\rangle\leftrightarrow|22\rangle$, see Appendix~\ref{IIIC}), intrinsic gate errors~\cite{Pedersen2007} can be suppressed below $10^{-4}$
with gate lengths $>$50 ns for both setups, see Figs.~\ref{fig3}(c)
and~\ref{fig3}(g). For high-coherence fluxoniums, the leading gate error could be from the relaxation
and dephasing of non-computational gate states, i.e., $|02(20)\rangle$ and $|12(21)\rangle$~\cite{Ficheux2021,Ding2023,Abad2023,Didier2019}.
Assuming the typical coherent times of $T_{1}^{21}=T_{2}^{21}=10\,\mu \rm s$~\cite{Ficheux2021,Ding2023}, we estimate that
for 50ns-CZ gates, the error ($\sim10^{-3}$) is limited by the decoherence of non-computational gate
states, while for shorter gates, the error
is dominated by leakage into non-computational states (see Appendix~\ref{IIIE}), see Figs.~\ref{fig3}(c) and~\ref{fig3}(g).

To assess the scalability of the architecture, Figs.~\ref{fig3}(d) and~\ref{fig3}(h) present
the 50ns-CZ gate error, accounting for the coupling between $Q_{1}$ and the spectator
$Q_{3}$, see Fig.~\ref{fig1}(a). The coupling strength, controlled by the coupler flux bias, is
quantified by the gate frequency shift $\delta\omega_{S}=|\omega_{11\rightarrow 21}^{(0)}-\omega_{11\rightarrow 21}^{(1)}|$,
where $\omega_{11\rightarrow 21}^{(0/1)}$ represent the gate frequency with $Q_{3}$ in $|0/1\rangle$. Following typical experimental
protocols~\cite{Krinner2020,Cai2021}, the CZ gate between $Q_{1}$ and $Q_{2}$ is tuned with $Q_{3}$ in $|0\rangle$ and then characterized
for both $|0\rangle$ and $|1\rangle$ states. As expected, increasing the $Q_1$-$Q_3$ coupling (see grey dots) generally raises
the gate error for both $Q_3$ states, with a more pronounced degradation for $|1\rangle$ (consistent with the
rise in $\delta\omega_{S}$), see Figs.~\ref{fig3}(d) and~\ref{fig3}(h). Notably, suppressing the qubit-spectator coupling
(e.g., $\delta\omega_{S}<100\,\rm kHz$) yields gate fidelity comparable to that in an isolated two-qubit
system with minimal spectator-state dependence. In contrast, strong coupling induces frequency collisions (see Appendix~\ref{IIID}) and
conditional frequency shifts that substantially degrade gate performance and introduce spectator-dependent correlated
errors. These analyses demonstrate that tunable interactions play a pivotal role in
mitigating spectator-related errors, thereby validating the scalability of our approach.

\section{Residual couplings with fluxonium biasing away from the half-flux quantum}\label{SecIV}

\begin{figure}[tbp]
\begin{center}
\includegraphics[keepaspectratio=true,width=\columnwidth]{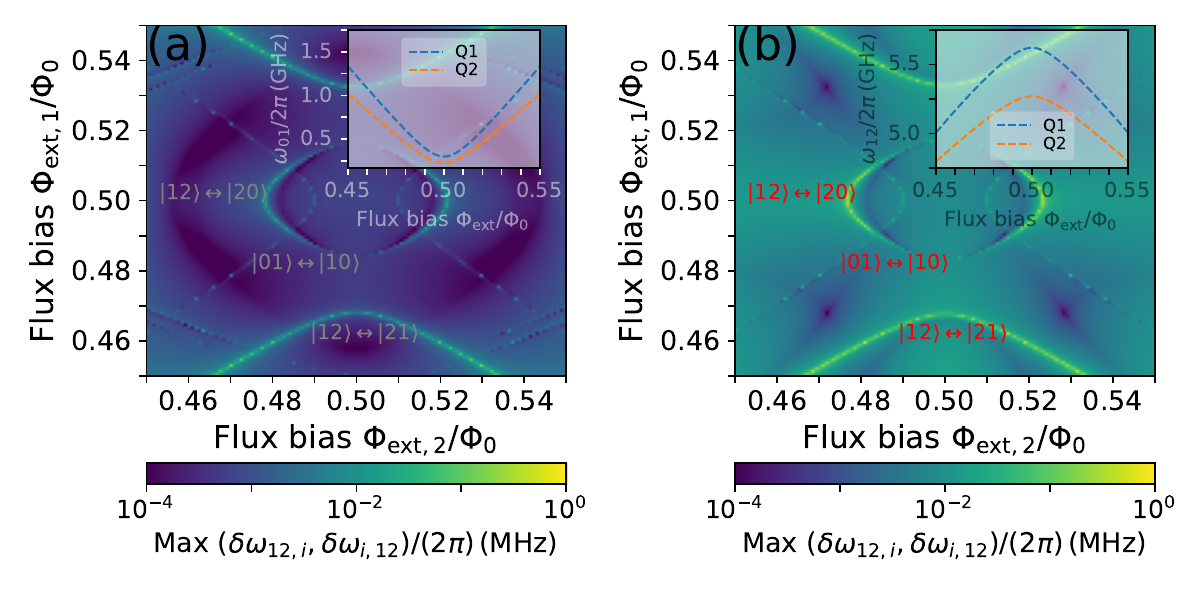}
\end{center}
\caption{Residual couplings, quantified by the maximum shifts ${\rm Max}(\delta\omega_{12,i},\delta\omega_{i,12}$), versus qubit flux bias. (a) and (b) are
for the DTC and STC setup, respectively. Sudden increases in data result from the frequency collisions, see grey
and red labels. Insets display the fluxonium frequency dependence
on the qubit flux bias.}
\label{fig4}
\end{figure}

While fluxonium qubits are typically biased at the half-flux sweet spot, certain applications, such as readout~\cite{Stefanski2024}, initialization~\cite{Manucharyan2009b,Zhang2021,Wang2024a}, and avoiding resonance with defects~\cite{Sun2023}, require
deviation from this point. Accordingly, plasmon interactions must also
be suppressed to prevent spectator-induced errors. Figure~\ref{fig4} shows that
the shifts can become significant when operating away from the sweet spot (re-tuning coupler can mitigate
residuals, but at the cost of increased tune-up complexity in scalable systems), particularly when frequency collisions happen.
Additionally, frequency collisions with coupler modes can also significantly degrade the residual suppression, see Appendix~\ref{I}.
These observations highlight the necessity of frequency allocation for
eliminating residual couplings.

\section{conclusion}\label{SecV}

In summary, this work identifies and addresses a critical challenge in scaling up fluxonium-based quantum processors
by proposing scalable coupling architectures to mitigate the issues of always-on plasmon interaction. By leveraging the
rich level structure with strong anharmonicity of fluxoniums, our proposed architecture combines near-complete decoupling of qubit states with
tunable plasmon interactions, effectively mitigating spectator-induced crosstalk and frequency collisions in both
computational and non-computational subspaces while preserving high-coherence in computational subspace and
enabling fast, high-fidelity CZ gates. Our comparative analysis of two possible implementations
elucidates the general principle for realizing the architectures while understanding and
addressing challenges with different implementations. We thus emphasize that the full potential of our results
lies not just in the advantage of one specific implementation, but rather the indication that scalable
architectures should address the issue due to always-on plasmon interactions and this can be achieved
by exploiting fluxonium's rich level structures and engineering tunable plasmon
interactions. Thus, we expect that our findings will also contribute to guide future scalable coupling designs
for quantum processors based on fluxoniums or other noise-protected superconducting qubits~\cite{Gyenis2021}.

\begin{acknowledgments}
Peng Zhao would like to thank Guoqiang Wang and Zhikun Han for the insightful discussions and also thank
Tao Jiang, Jianbin Cai, Naibin Zhou, Rui Wang, Tan He, Weiping Lin, Dongxin Gao, Teng Ma, and Peng Xu for their generous support.
This work is supported by the National Natural Science Foundation of China (Grants No.12204050), the Innovation
Program for Quantum Science and Technology (Grant No. 2021ZD0300200), Shanghai Municipal Science and Technology
Major Project (Grant No.2019SHZDZX01), Anhui Initiative in Quantum Information Technologies, and the Special funds
from Jinan science and Technology Bureau and Jinan high tech Zone Management Committee.

Peng Zhao and Guming Zhao contributed equally to this work.
\end{acknowledgments}


\appendix

\makeatletter
\renewcommand\thetable{S\@arabic\c@table}
\renewcommand \thefigure{S\@arabic\c@figure}
\renewcommand \theequation{S\@arabic\c@equation}
\makeatother

\section{The fluxonium architecture with double-transmon couplers}\label{I}

\begin{figure}[htbp]
\begin{center}
\includegraphics[keepaspectratio=true,width=\columnwidth]{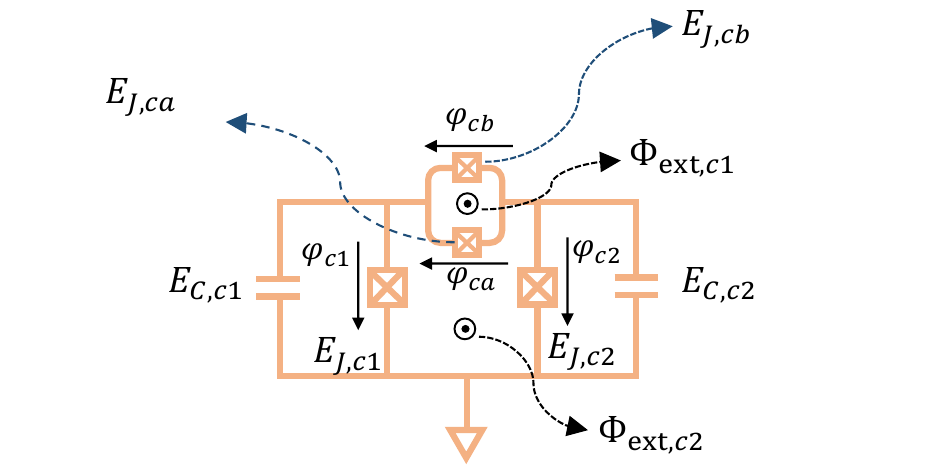}
\end{center}
\caption{Coupler circuit. The coupler comprises two transmons coupled inductively via a dc SQUID. The strength
of the intermode inductive coupling can be controlled via the flux bias applied to the main
loop ($\Phi_{{\rm ext},c2}$) or the SQUID loop ($\Phi_{{\rm ext},c1}$).}
\label{fig5}
\end{figure}

Here, we begin by presenting detailed derivations of both the full system Hamiltonian and the effective Hamiltonian
for the architecture based on double-transmon couplers (DTCs), as shown in Fig.~\ref{fig5}, and then study the impact of
various imperfections including stray capacitive couplings and variations in Josephson Junctions on
the functionality of the DTC-based implementation. Finally, we analyze residual couplings
with different coupler parameters and evaluate the coupler-induced qubit decoherence.
Hereafter, to describe the system state within the DTC setup, we use the notation
of $|Q_{1}Q_{2},C_{1}C_{2}\rangle$ and when confined to qubit subspace,
notation $|Q_{1}Q_{2}\rangle$ is used for $|Q_{1}Q_{2},00\rangle$.

\begin{figure*}[htbp]
\begin{center}
\includegraphics[width=16cm,height=7cm]{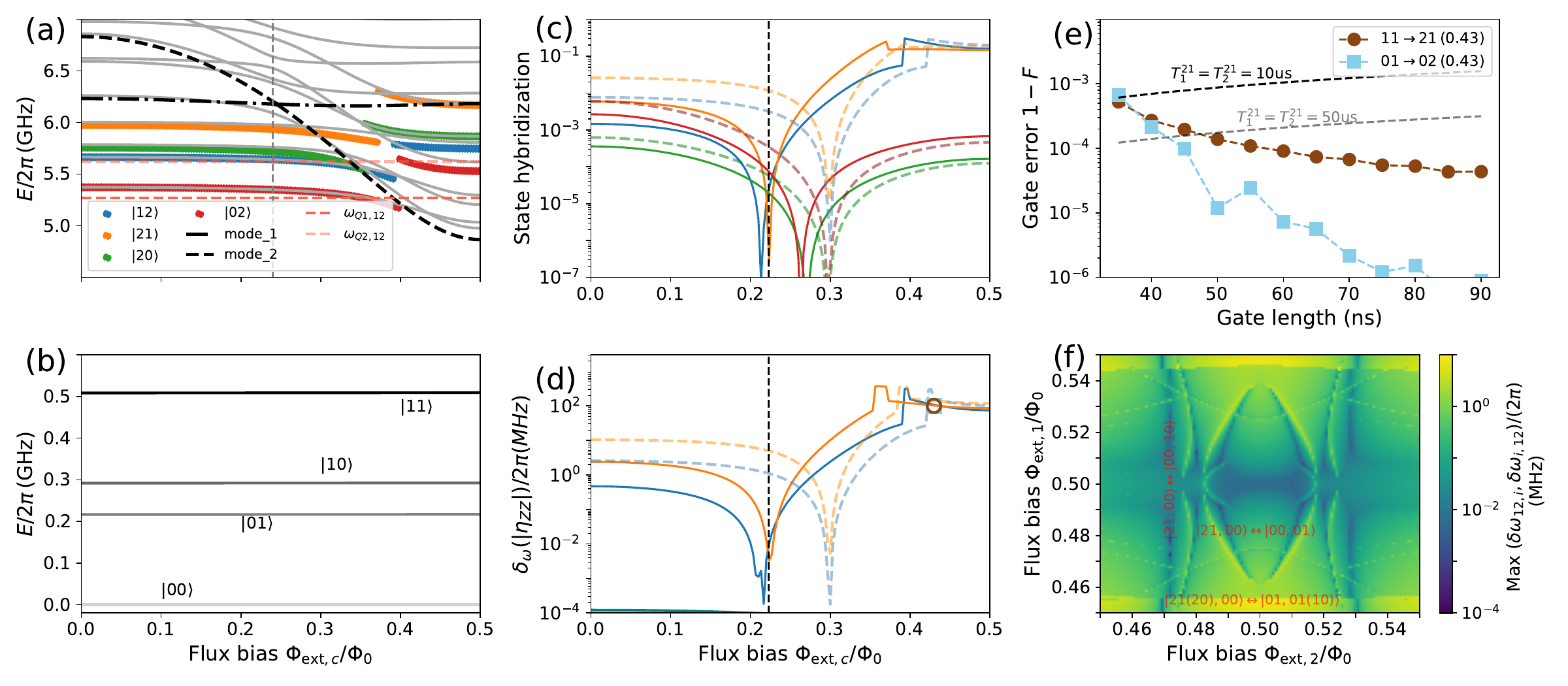}
\end{center}
\caption{Main results for the Type-2 setup with qubit parameters same as in the Type-1 setup and coupler parameters
summarized in Table~\ref{tab:setup-2}. (a,b) The system spectrum versus the coupler flux bias. Here, the coupler's eigenmode frequencies and
the fluxonium's bare plasmon frequencies are also presented with vertical grey dashed lines denoting the bias for nulling
the DTC's intermode coupling. (c,d) Coupler-mediated plasmon interactions, quantified
by state hybridization (c) and state-dependent plasmon frequency shifts (d), versus the coupler flux bias (sudden jumps
or discontinuities in curves are caused by state labeling failure near avoided crossings). Here, vertical black dashed
lines denote the bias for minimizing the frequency shift. (e) CZ gate errors with
varying gate lengths. (f) Residual couplings, quantified by the maximum shifts ${\rm Max}(\delta\omega_{12,i},\delta\omega_{i,12}$),
versus qubit flux bias. Sudden increases or discontinuities in shift result from the frequency collisions involving fluxonium
transitions or the coupler modes (as indicated by the red labels). We note that in (c,d), sudden jumps or discontinuities in curves
are caused by state labeling failure near avoided crossings.}
\label{setup_V2}
\end{figure*}

In the main text, we study the DTC-based implementation, where the coupler is controlled by the flux bias
applied only to the SQUID loop (hereafter, nicknamed Type-1 setup). Here, for comparative analysis, we
additionally investigate the Type-2 setup, where the flux bias is applied to the main loop, and noting that
practical realizations can substitute the dc SQUID with a single Josephson junction.
Following the same analytical framework applied to Type-1 in the main text, Figure~\ref{setup_V2} presents the
corresponding results for the Type-2 setup with qubit parameters same as in the Type-1 setup (see Table~\ref{tab:parameters} of the
main text) and coupler parameters summarized in Table~\ref{tab:setup-2}.

\begin{table}[!htb]
\caption{\label{tab:setup-2} Coupler parameters for the Type-2 setup.}
\begin{ruledtabular}
\begin{tabular}{cccc}
$ $&
$E_C$ (GHz)&
$E_L$ (GHz)&
$E_J$ (GHz)\\\hline
Transmon $ci$ & 0.25 & $-$ & 20 \\
\hline
\hline
$ $ & $J_{c1 (c2)}$ (MHz) & $J_{C,12}$ (MHz)  & $E_{J,12}$ (GHz)\\\hline
Coupling strengths & 550  & 100   & 3.5
\end{tabular}
\end{ruledtabular}
\end{table}

\subsection{Coupler Circuit}\label{IA}

Figure~\ref{fig5} depicts the two-mode coupler, which comprises two transmons coupled inductively via
a dc SQUID. Here, $E_C$ and $E_J$  denote the charging and Josephson energies, respectively, the phase
difference across the Josephson junction $ci$ is $\varphi_{ci}$, and the subscripts $c1\,(c2)$ and $ca\,(cb)$
correspond to the transmon and the dc SQUID. The potential energy of the coupler is (assuming
that $E_{J,ca}>E_{J,cb}$)
\begin{equation}
\begin{aligned}\label{eqA1}
U_{c}=\sum_{i=1,2}[-E_{J,ci}\cos\varphi_{ci}]-(E_{J,ca}\cos\varphi_{ca}+E_{J,cb}\cos\varphi_{cb}).
\end{aligned}
\end{equation}
The fluxoid quantization in the main and SQUID loops leads to the following two relations,
i.e., $\varphi_{c1}-\varphi_{c2}+\varphi_{ca}+\varphi_{\text{ext},c2}=0$ and
$\varphi_{cb}-\varphi_{ca}+\varphi_{\text{ext},c1}=0$ ($\varphi_\text{ext}=2\pi\Phi_\text{ext}/\Phi_0$),
giving rise to $\varphi_{ca}=-(\varphi_{c1}-\varphi_{c2})-\varphi_{\text{ext},c2}$
and $\varphi_{cb}=\varphi_{ca}-\varphi_{\text{ext},c1}$. Accordingly, the coupler potential
energy can be rewritten as
\begin{equation}
\begin{aligned}\label{eqA2}
U_{c}&=-\sum_{i=1,2}[E_{J,ci}\cos\varphi_{ci}]\\
&\quad -E_{J,ca}[\cos\varphi_{ca}+E_{J,cb}\cos(\varphi_{ca}-\varphi_{\text{ext},c1})]\\
&=-\sum_{i=1,2}[E_{J,ci}\cos\varphi_{ci}]\\
&\quad -E_{J,12}\cos(\varphi_{c1}-\varphi_{c2}+\varphi_{\text{ext},c2}+\varphi_{0}),
\end{aligned}
\end{equation}
with
\begin{equation}
\begin{aligned}\label{eqA3}
&E_{J,12}=E_{J,c12}\cos\left(\frac{\varphi_{\text{ext},c1}}{2}\right)\sqrt{1+d^{2}\tan^{2}\frac{\varphi_{\text{ext},c1}}{2}},
\\&\varphi_{0}=\arctan\left[\frac{(1-d)\tan\frac{\varphi_{\text{ext},c1}}{2}}{1+d\tan^{2}\frac{\varphi_{\text{ext},c1}}{2}}\right],
\end{aligned}
\end{equation}
where $E_{J,c12}=E_{J,ca}+E_{J,cb}$ is the sum of the two Josephson energies of the SQUID
and $d=(E_{J,ca}-E_{J,cb})/(E_{J,ca}+E_{J,cb})$ quantifies the junction asymmetries in the
SQUID. When the SQUID's junctions are identical, i.e., $E_{J,ca}=E_{J,cb}$,
the coupler potential energy is reduced to the well-known form ~\cite{Orlando1999,Paauw2009}
\begin{equation}
\begin{aligned}\label{eqA4}
U_{c}=&-\sum_{i=1,2}[E_{J,ci}\cos\varphi_{ci}]-E_{J,c12}\cos\frac{\varphi_{\text{ext},c1}}{2}\\
&\times\left[\cos(\varphi_{c1}-\varphi_{c2}+\varphi_{\text{ext},c2}+\frac{\varphi_{\text{ext},c1}}{2})\right],
\end{aligned}
\end{equation}
which clearly reveals the intrinsic flux crosstalk between the biases applied to the SQUID and
the main loops.

As our study only involves the low-lying energy states of the coupler and we assume
that the coupler modes are operated in the transmon regime (giving small phase fluctuations around the
low-lying energy states)~\cite{Koch2007}, the potential Hamiltonian of the coupler can be approximated
by \cite{Miyanaga2021,Goto2022,Campbell2023,Geller2015}
\begin{equation}
\begin{aligned}\label{eqA5}
U_{c}\approx &\sum_{i=1,2}\frac{\hat\varphi_{ci}^{2}}{2}[E_{J,ci}\cos\bar\varphi_{ci}\\
&+E_{J,12}\cos(\bar\varphi_{c1}-\bar\varphi_{c2}+\varphi_{\text{ext},c2}+\varphi_{0})]\\
&-E_{J,12}\cos(\bar\varphi_{c1}-\bar\varphi_{c2}+\varphi_{\text{ext},c2}+\varphi_{0})\hat\varphi_{c1}\hat\varphi_{c2}
\end{aligned}
\end{equation}
where $\bar\varphi_{c1}$ and $\bar\varphi_{c2}$ are determined to minimize the potential energy $U_{c}$ and satisfy the
following two conditions, i.e., the first-order derivation of the potential energy over $\varphi_{ci}$ should vanish,
\begin{equation}
\begin{aligned}\label{eqA6}
&\sin\bar\varphi_{c1}=-\frac{E_{J,12}}{E_{J,c1}}\sin(\bar\varphi_{c1}-\bar\varphi_{c2}+\varphi_{\text{ext},c2}+\varphi_{0}),\\
&\sin\bar\varphi_{c2}=\frac{E_{J,12}}{E_{J,c2}}\sin(\bar\varphi_{c1}-\bar\varphi_{c2}+\varphi_{\text{ext},c2}+\varphi_{0}).
\end{aligned}
\end{equation}

For the Type-1 setup, we have $\varphi_{\text{ext},c2}+\varphi_{0}=0$ and
thus $\bar\varphi_{ci}=0$ (for clarity, here we assume that $d=0$ and the intrinsic
flux crosstalk between the main and SQUID loops are compensated for solely biasing
the SQUID loop), while for the Type-2 setup, where the dc SQUID is replaced with
a single junction, we thus have $\varphi_{\text{ext},c1}=0$. Consequently, the following approximation of the
potential Hamiltonian can be obtained for the two operational setups, i.e.,
\begin{equation}
\begin{aligned}\label{eqA7}
&U_{c}^{(1)}\approx \sum_{i=1,2}\frac{\hat\varphi_{ci}^{2}}{2}[E_{J,ci}+E_{J,c12}\cos\frac{\varphi_{\text{ext},c1}}{2}]\\
&\quad\quad -E_{J,c12}\cos\frac{\varphi_{\text{ext},c1}}{2}\hat\varphi_{c1}\hat\varphi_{c2},\\
&U_{c}^{(2)}\approx \sum_{i=1,2}\frac{\hat\varphi_{ci}^{2}}{2}[E_{J,ci}+E_{J,c12}\cos(\bar\varphi_{c1}-\bar\varphi_{c2}+\varphi_{\text{ext},c2})]\\
&\quad\quad -E_{J,c12}\cos(\bar\varphi_{c1}-\bar\varphi_{c2}+\varphi_{\text{ext},c2})\hat\varphi_{c1}\hat\varphi_{c2},
\end{aligned}
\end{equation}
where the second term of the first line of $U_{c}$ corresponds to the correction to the transmon's potential
energy arising from the intermode inductive coupling and the second line describes the dipole-dipole
interaction between the two transmon modes.

Accordingly, the intermode inductive coupling Hamiltonians for the two setups are
\begin{equation}
\begin{aligned}\label{eqA8}
&H_{c}^{(1)}= -E_{J,c12}\cos\frac{\varphi_{\text{ext},c1}}{2}\hat\varphi_{c1}\hat\varphi_{c2},\\
&H_{c}^{(2)}=-E_{J,c12}\cos(\bar\varphi_{c1}-\bar\varphi_{c2}+\varphi_{\text{ext},c2})\hat\varphi_{c1}\hat\varphi_{c2},\\
&\quad\quad =-E_{J,c12}\cos(\tilde\varphi_{\text{ext},c2})\hat\varphi_{c1}\hat\varphi_{c2},
\end{aligned}
\end{equation}
where $\tilde\varphi_{\text{ext},c2}$ denotes the effective external phase bias for the coupler and according
to Eq.~(\ref{eqA6}), its relation to the coupler phase bias $\varphi_{\text{ext},c2}$ is described by
\begin{equation}
\begin{aligned}\label{eqA9}
&\varphi_{\text{ext},c2}=\tilde\varphi_{\text{ext},c2}+\arcsin\left(\frac{E_{J,c12}}{E_{J,c1}}\sin\tilde\varphi_{\text{ext},c2}\right)\\
&\quad\quad+\arcsin\left(\frac{E_{J,c12}}{E_{J,c2}}\sin\tilde\varphi_{\text{ext},c2}\right).
\end{aligned}
\end{equation}

After including kinetic energies of the coupler circuit, see Fig.~\ref{fig5}, the full Hamiltonian of the
coupler for the Type-1 setup is
\begin{equation}
\begin{aligned}\label{eqA10}
H_{\rm coupler}^{(1)}\approx &\sum_{i=1,2} \left[4 E_{C,ci} \hat n^2_{ci} + (E_{J,ci}+E_{J,c12}\cos\frac{\varphi_{\text{ext},c}}{2})\hat\varphi_{ci}^{2}\right]\\
&-E_{J,c12}\cos\frac{\varphi_{\text{ext},c}}{2}\hat\varphi_{c1}\hat\varphi_{c2},
\end{aligned}
\end{equation}
and for the Type-2 setup, the coupler Hamiltonian is
\begin{equation}
\begin{aligned}\label{eqA11}
H_{\rm coupler}^{(2)}\approx &\sum_{i=1,2} \left[4 E_{C,ci} \hat n^2_{ci} + (E_{J,ci}+E_{J,c12}\cos\tilde{\varphi}_{\text{ext},c})\hat\varphi_{ci}^{2}\right]\\
&-E_{J,c12}\cos\tilde{\varphi}_{\text{ext},c}\hat\varphi_{c1}\hat\varphi_{c2}.
\end{aligned}
\end{equation}
We note here that although Equations~(\ref{eqA10}) and~(\ref{eqA11}) have the same form, their physical
implications differ significantly at intermode interaction nulling points, i.e.,
$\varphi_{{\rm ext},c}=\pi$ and $\tilde{\varphi}_{{\rm ext},c2}=\pi/2$ for the two setups. In
the Type-1 setup, the intermode inductive coupling is rigorously nulled regardless of approximation, see
Eqs.~(\ref{eqA4}) and~(\ref{eqA10}), whereas in the Type-2 setup, this cancelation is only approximate
due to truncation of higher-order terms in the derivation of Eq.~(\ref{eqA11}).

By approximating transmon modes as linear harmonic modes, i.e., neglecting nonlinear parts
of the transmon (replacing transmons with linear oscillators, see~\cite{Miyanaga2021}), and expressing the phase and number operator as~\cite{Koch2007}
\begin{equation}
\begin{aligned}\label{eqA12}
&\hat \varphi_{ci} = \phi_{ci,{\rm zpf}}(\hat a_{ci}^{\dag}+\hat a_{ci}),\quad \hat n_{ci} = i n_{ci,{\rm zpf}}(\hat a_{ci}^{\dag}-\hat a_{ci}),\\
&\varphi_{ci,{\rm zpf}} =\frac{1}{\sqrt{2}}\left(\frac{8E_{C,ci}}{E_{J,ci}}\right)^{\frac{1}{4}},\quad n_{ci,{\rm zpf}}= \frac{1}{\sqrt{2}}\left(\frac{E_{J,ci}}{8E_{C,ci}}\right)^{\frac{1}{4}},
\end{aligned}
\end{equation}
where $a_{ci}$ ($a_{ci}^{\dag}$) denotes the destroy (creation) operator for the coupler
mode of $\omega_{ci}$ and $\phi_{ci,{\rm zpf}}$ ($n_{ci,{\rm zpf}}$) represents the
phase (number) zero-point fluctuation, the coupler Hamiltonian can be approximated by
\begin{equation}
\begin{aligned}\label{eqA13}
H_{\rm coupler}=\sum_{i=1,2}[\omega_{ci}\hat a_{ci}^{\dag}\hat a_{ci}]
+g_{c}(\hat a_{c1}+\hat a_{c1}^{\dag})(\hat a_{c2}+\hat a_{c2}^{\dag}),
\end{aligned}
\end{equation}
where $g_{c}=-g_{c,{\rm ind}}$ denotes the intermode coupling strength with $g_{c,{\rm ind}}=E_{J,12}\cos(\varphi_{\text{ext},c}/2)\varphi_{c1,{\rm zpf}}\varphi_{c2,{\rm zpf}}$ and $g_{c,{\rm ind}}=E_{J,12}\cos(\tilde{\varphi}_{\text{ext},c})\varphi_{c1,{\rm zpf}}\varphi_{c2,{\rm zpf}}$
for the two setups, respectively, see Eq.~(\ref{eqA8}).

\subsection{Coupler-mediated effective couplings}\label{IB}

Here, we present the derivation of coupler-mediated effective interactions for a selected
fluxonium transition $|k\rangle\leftrightarrow|l\rangle$ of $\omega_{kl,i}$. The effective Hamiltonian is derived
both with and without application of the rotating-wave approximation (RWA) to the intermode
coupling Hamiltonian, see Eq.~(\ref{eqA13}). Both derivations show that the coupler-mediated
fluxonium interactions can be programmable via the coupler flux bias (i.e., tuning the intermode
coupling strength) and the interaction nulling conditions is only dependent on the intermode
coupling while is relevant to fluxonium parameters.

In the following discussion, for clarity, we assume that the two coupler modes are
degenerate, i.e., $\omega_{c1(c2)}=\omega_{c}$, and omit non-RWA terms in the
fluxonium-coupler interaction Hamiltonian, whose contribution to the effective coupling
strength scales as $1/(\omega_{kl,i}+\omega_{c})$ \cite{Yan2018}. To derive the effective
Hamiltonian, we further assume that the fluxonium-coupler interaction strength $g_{kl,i}$ is far smaller than
the fluxonium-coupler detuning $\Delta_{kl,i}=\omega_{kl,i}-\omega_{c}$ (i.e., the
dispersive regime) and $g_{c}\ll\omega_{c}$, especially for the system biased around the interaction
nulling point (where the coupler's intermode coupling is turned off). However, we emphasize that to
mediate strong plasmon interactions for implementing fast CZ
gates, the dispersive condition generally breaks down around the coupler interaction points.

\subsubsection{With the RWA}\label{IB1}

After applying the RWA, the full system Hamiltonian is
\begin{equation}
\begin{aligned}\label{eqA14}
H_{kl}^{(D)}=& \sum_{i=1,2} [\omega_{kl,i}\sigma_{kl,i}^{\dag}\sigma_{kl,i}+g_{kl,i}(\sigma_{kl,i}^{\dag}a_{ci}+\sigma_{kl,i}a_{ci}^{\dag})]\\
&+\sum_{i=1,2}[\omega_{c}a_{ci}^{\dag}a_{ci}]+g_{c}(a_{c1}^{\dag}a_{c2}+a_{c1}a_{c2}^{\dag}),
\end{aligned}
\end{equation}
where $\hat\sigma_{kl,i}=|k\rangle\langle l|_{i}$ ($\sigma_{kl,i}^{\dag}$) is the lowering (raising)
operator for the transition of $\omega_{kl,i}$ and $g_{kl,i}=J_{ci} \langle k1|\hat n_i \hat n_{ci}|l0\rangle$
represents the fluxonium-coupler coupling strength. The effective fluxonium interactions can be obtained by
eliminating the direct fluxonium-coupler interactions. This can be achieved by firstly diagonalizing the
coupler Hamiltonian and then eliminating the direct interaction between the fluxoniums and the
coupler eigenmodes.

After diagonalizing the two coupled degenerate modes, the coupler Hamiltonian
of Eq.~(\ref{eqA13}) becomes
\begin{equation}
\begin{aligned}\label{eqA15}
H_{\rm coupler}^{\rm RWA}&= \sum_{i=1,2}[\omega_{c}\hat a_{ci}^{\dag}\hat a_{ci}]
+g_{c}(\hat a_{c1}^{\dag}\hat a_{c2}+\hat a_{c1}\hat a_{c2}^{\dag})
\\&=\omega_{+}\hat a_{+}^{\dag}\hat a_{+}+\omega_{-}\hat a_{-}^{\dag}\hat a_{-},
\end{aligned}
\end{equation}
where $\omega_{\pm}=\omega_{c}\pm g_{c}$ ($g_{c}=-g_{c,{\rm ind}}$) denotes the eigenmode frequency
of the coupler and $\hat a_{\pm}$ ($\hat a_{\pm}^{\dag}$) corresponds to the destroy (creation) operator for
the eigenmodes with
\begin{equation}
\begin{aligned}\label{eqA16}
\hat a_{\pm}=\frac{\hat a_{c1}\pm \hat a_{c2}}{\sqrt{2}}, \hat a_{\pm}^{\dag}=\frac{\hat a_{c1}^{\dag}\pm \hat a_{c2}^{\dag}}{\sqrt{2}}.
\end{aligned}
\end{equation}
Accordingly, by inserting
\begin{equation}
\begin{aligned}\label{eqA17}
\hat a_{c1}=\frac{\hat a_{+}+ \hat a_{-}}{\sqrt{2}}, \hat a_{c2}=\frac{\hat a_{+}-\hat a_{-}}{\sqrt{2}},
\end{aligned}
\end{equation}
into the full system Hamiltonian of Eq.~(\ref{eqA14}), we have
\begin{equation}
\begin{aligned}\label{eqA18}
H_{kl}^{(D)}=& \sum_{i=1,2} [\omega_{kl,i}\hat\sigma_{kl,i}^{\dag}\hat\sigma_{kl,i}]+\omega_{+}\hat a_{+}^{\dag}\hat a_{+}+\omega_{-}\hat a_{-}^{\dag}\hat a_{-}
\\&+\frac{g_{kl,1}}{\sqrt{2}}(\hat\sigma_{kl,1}^{\dag}\hat a_{+}+h.c.)+\frac{g_{kl,2}}{\sqrt{2}}(\hat\sigma_{kl,2}^{\dag}\hat a_{+}+h.c.)
\\&+\frac{g_{kl,1}}{\sqrt{2}}(\hat \sigma_{kl,1}^{\dag}\hat a_{-}+h.c.)-\frac{g_{kl,2}}{\sqrt{2}}(\hat\sigma_{kl,2}^{\dag}\hat a_{-}+h.c.).
\end{aligned}
\end{equation}
Considering that the fluxonium-coupler system are operated in the dispersive regime, an effective Hamiltonian
can be obtained by removing the two eigenmodes \cite{Blais2021}, i.e.,
\begin{equation}
\begin{aligned}\label{eqA19}
H_{kl,{\rm eff}}^{(D)}\approx& \omega_{kl,1}\hat\sigma_{kl,1}^{\dag}\hat\sigma_{kl,1}+\omega_{kl,2}\hat\sigma_{kl,2}^{\dag}\hat\sigma_{kl,2}
\\&+g_{kl,{\rm eff}}(\hat\sigma_{kl,1}^{\dag}\hat\sigma_{kl,2}+\hat\sigma_{kl,1}\hat\sigma_{kl,2}^{\dag}),
\end{aligned}
\end{equation}
where $g_{kl,{\rm eff}}$ denotes the strength of the effective couplings. Taking the first-order approximation, i.e., keeping
terms to the order of $g_{c}/\omega_{c}$, the effective coupling strength is given as~\cite{McKay2015}
\begin{equation}
\begin{aligned}\label{eqA20}
g_{kl,{\rm eff}}^{(D)}&=\frac{1}{2}\frac{g_{kl,1}}{\sqrt{2}}\frac{g_{kl,2}}{\sqrt{2}}
\left(\frac{1}{\omega_{kl,1}-\omega_{+}}+\frac{1}{\omega_{kl,2}-\omega_{+}}\right) \\
&\quad -\frac{1}{2}\frac{g_{kl,1}}{\sqrt{2}}\frac{g_{kl,2}}{\sqrt{2}}\left(\frac{1}{\omega_{kl,1}-\omega_{-}}+\frac{1}{\omega_{kl,2}-\omega_{-}}\right)\\
&\approx\frac{g_{kl,1}g_{kl,2}g_{c}}{2}\left(\frac{1}{\Delta_{kl,1}^{2}}+\frac{1}{\Delta_{kl,2}^{2}}\right).
\end{aligned}
\end{equation}

\subsubsection{Without the RWA}\label{IB2}

Without applying the RWA to the coupler Hamiltonian in Eq.~(\ref{eqA13}), the full system Hamiltonian can
be described by (the same as the Eq.~(\ref{eq2}) of the main text)
\begin{equation}
\begin{aligned}\label{eqA21}
H_{kl}^{(D)}=& \sum_{i=1,2} [\omega_{kl,i}\hat\sigma_{kl,i}^{\dag}\hat\sigma_{kl,i}+g_{kl,i}(\hat\sigma_{kl,i}^{\dag}\hat a_{ci}+\hat\sigma_{kl,i}\hat a_{ci}^{\dag})]\\
&+\sum_{i=1,2}[\omega_{c}\hat a_{ci}^{\dag}\hat a_{ci}]+g_{c}(\hat a_{c1}+\hat a_{c1}^{\dag})(\hat a_{c2}+\hat a_{c2}^{\dag}),
\end{aligned}
\end{equation}
Similar to the above derivation with RWA, the effective Hamiltonian is obtained by firstly diagonalizing the
coupler Hamiltonian. Here, the coupler Hamiltonian can be diagonalized by employing the Bogoliubov transformation
$[\hat a_{c1},\hat a_{c1}^{\dag},\hat a_{c2},\hat a_{c2}^{\dag}]^{T}=M [\hat a_{-},\hat a_{-}^{\dag},\hat a_{+},\hat a_{+}^{\dag}]^{T}$~\cite{Zhang2023}.
Keeping terms to the order of $g_{c}/\omega_{c}$, the transformation matrix $M$ is expressed by
\begin{equation}\label{eqA22}
M\approx\left[
\begin{array}{cccc}
\frac{1}{\sqrt{2}} & \frac{g_{c}}{2\sqrt{2}\omega_{c}} & \frac{1}{\sqrt{2}} & -\frac{g_{c}}{2\sqrt{2}\omega_{c}} \\
\frac{g_{c}}{2\sqrt{2}\omega_{c}} & \frac{1}{\sqrt{2}} & -\frac{g_{c}}{2\sqrt{2}\omega_{c}} & \frac{1}{\sqrt{2}} \\
-\frac{1}{\sqrt{2}} & -\frac{g_{c}}{2\sqrt{2}\omega_{c}} & \frac{1}{\sqrt{2}} & -\frac{g_{c}}{2\sqrt{2}\omega_{c}}\\
-\frac{g_{c}}{2\sqrt{2}\omega_{c}} & -\frac{1}{\sqrt{2}} & -\frac{g_{c}}{2\sqrt{2}\omega_{c}} & \frac{1}{\sqrt{2}}\\
\end{array}
\right].
\end{equation}
and the eigenmode frequency of the coupler is $\omega_{\pm}=\sqrt{\omega_{c}^{2}\pm2g_{c}\omega_{c}}\approx\omega_{c}\pm g_{c}$.

According to the the transformation matrix $M$ in Eq.~(\ref{eqA22}), the interaction Hamiltonian
of the coupled fluxonium system can be rewritten as
\begin{equation}
\begin{aligned}\label{eqA23}
H_{kl,I}^{(D)}&=\sum_{i=1,2}\left[g_{kl,i}(\hat\sigma_{kl,i}\hat a_{i}^{\dag}+h.c.)\right]\\
&=g_{kl,1}\left(\frac{1}{\sqrt{2}}-\frac{g_{c}}{2\sqrt{2}\omega_{c}}\right)(\hat a_{-}\hat\sigma_{kl,1}^{\dag}+h.c.)\\
&\quad +g_{kl,1}\left(\frac{1}{\sqrt{2}}+\frac{g_{c}}{2\sqrt{2}\omega_{c}}\right)(\hat a_{+}\hat\sigma_{kl,1}^{\dag}+h.c.)\\
&\quad -g_{kl,2}\left(\frac{1}{\sqrt{2}}-\frac{g_{c}}{2\sqrt{2}\omega_{c}}\right)(\hat a_{-}\hat\sigma_{kl,2}^{\dag}+h.c.)\\
&\quad +g_{kl,2}\left(\frac{1}{\sqrt{2}}+\frac{g_{c}}{2\sqrt{2}\omega_{c}}\right)(\hat a_{+}\hat\sigma_{kl,2}^{\dag}+h.c.).
\end{aligned}
\end{equation}
Again, removing the direct fluxonium-mode couplings gives rise to the effective interaction Hamiltonian $H_{kl,I}^{\rm eff}= g_{kl,{\rm eff}}(\hat\sigma_{kl,1}^{\dag}\hat\sigma_{kl,2}+\hat\sigma_{kl,1}\hat\sigma_{kl,2}^{\dag})$ with the effective coupling strength
\begin{equation}
\begin{aligned}\label{eqA24}
g_{kl,{\rm eff}}^{(D)}=&\frac{g_{kl,1}g_{kl,2}}{2}\left(\frac{1}{\sqrt{2}}+\frac{g_{c}}{2\sqrt{2}\omega_{c}}\right)^{2}
\sum_{i=1,2}\left(\frac{1}{\Delta_{kl,i+}}\right)\\
&-\frac{g_{kl,1}g_{kl,2}}{2}\left(\frac{1}{\sqrt{2}}-\frac{g_{c}}{2\sqrt{2}\omega_{c}}\right)^{2}
\sum_{i=1,2}\left(\frac{1}{\Delta_{kl,i-}}\right)\\
\end{aligned}
\end{equation}
where $\Delta_{kl,i\pm}=\omega_{kl,i}-\omega_{\pm}$ represents the detuning between the fluxonium frequency and
the coupler's eigenmode frequency. Keeping terms to the order of $g_{c}/\omega_{c}$, the coupling strength
is approximated by
\begin{equation}
\begin{aligned}\label{eqA25}
g_{kl,{\rm eff}}^{(D)}\approx\frac{g_{kl,1}g_{kl,2}g_{c}}{2}
\sum_{i=1,2}\left(\frac{1}{\Delta_{kl,i}^{2}}+\frac{\Delta_{kl,i}}{\omega_{c}}\frac{1}{\Delta_{kl,i}^{2}}\right),
\end{aligned}
\end{equation}
recovering the formula shown in Eq.~(\ref{eq4}) of the main text.

\subsection{Stray capacitive couplings}\label{IC}

In the above derivation, we neglect parasitic couplings due to stray capacitances
in the coupling circuit. However, for practical systems, stray capacitive couplings
are ubiquitous and can not only create nearest-neighbor interactions but
can also enable parasitic couplings beyond nearest neighbors. Crucially, they may significantly
degrade the coupler's on/off ratio and introduce substantial residual couplings. Here, we turn
to investigate the impact of these stray couplings on coupler performance.

\begin{figure}[htbp]
\begin{center}
\includegraphics[keepaspectratio=true,width=\columnwidth]{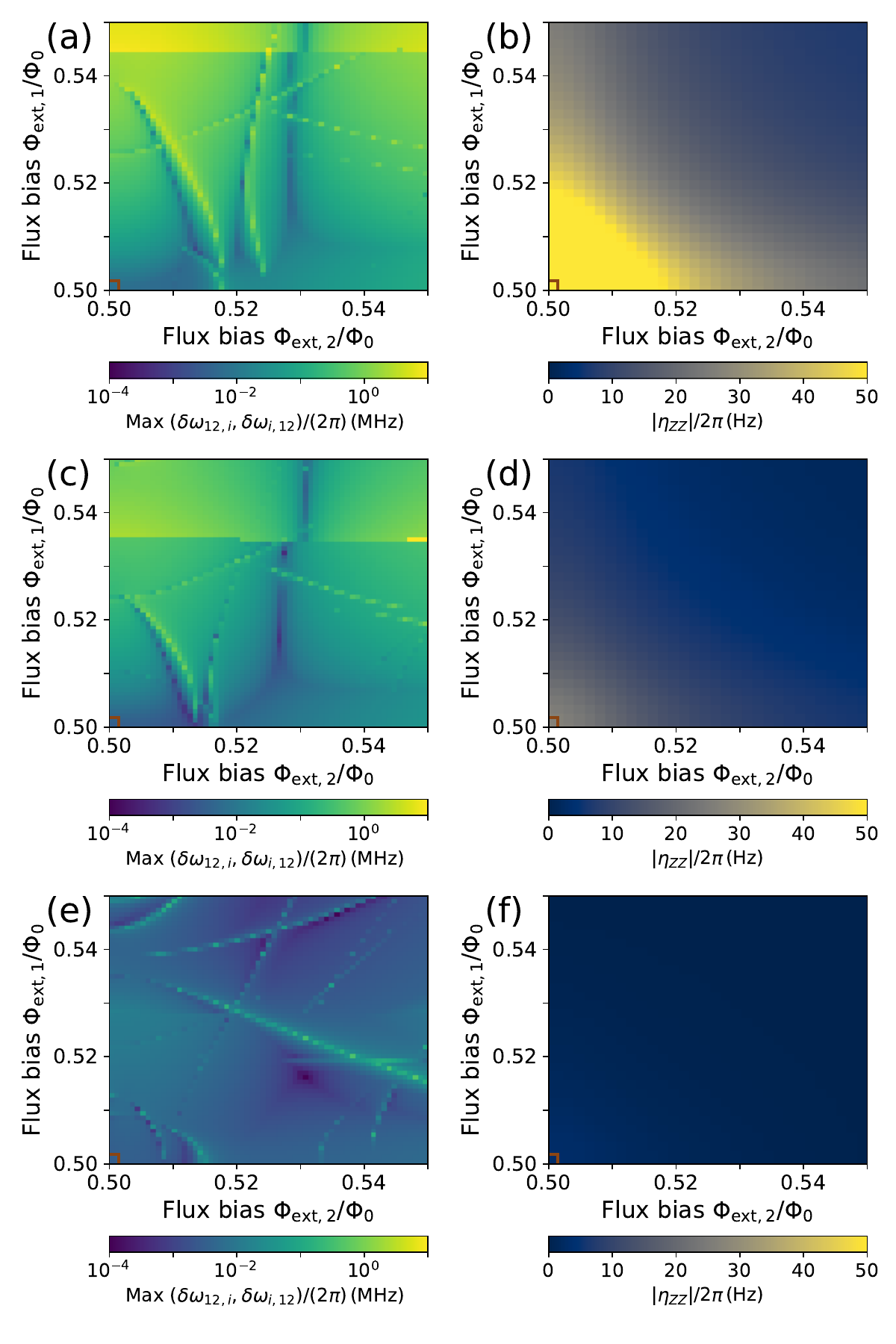}
\end{center}
\caption{Residual couplings, quantified by (left panels) the maximum state-dependent shifts ${\rm Max}(\delta\omega_{12,i},\delta\omega_{i,12}$)
and (right panels) the ZZ coupling strength $\eta_{ZZ}$, as functions of the qubit flux bias with different intermode capacitive
coupling strengths for the Type-2 setup. (a,b) $J_{C,12}/2\pi=100\,{\rm MHz}$, as the same as that in Fig.~\ref{setup_V2}(f).
(c,d) $J_{C,12}/2\pi=50\,{\rm MHz}$. (e,f) $J_{C,12}/2\pi=0\,{\rm MHz}$. Other system parameters are
the same as that used in Fig.~\ref{setup_V2}.}
\label{fig6}
\end{figure}
\subsubsection{Intermode capacitive couplings}\label{IC1}

In the current coupler design, see Fig.~\ref{fig1}(b) of the main text, the dominated parasitic
neighboring couplings comes from the self-capacitance of the coupling junction (typically a few fFs~\cite{Chow2015}), leading
to an additional intermode capacitive coupling for the coupler, i.e.,
\begin{equation}
\begin{aligned}\label{eqB1}
H_{c,\text{cap}}&=J_{C,12} \hat n_{c1} \hat n_{c2}\\
&=-g_{c,{\rm cap}}(\hat a_{c1}-\hat a_{c1}^{\dag})(\hat a_{c2}-\hat a_{c2}^{\dag}),
\end{aligned}
\end{equation}
where $g_{c,{\rm cap}}$ denotes the strength of the intermode capacitive coupling.

Similar to the derivation given in Sec.~\ref{IB1}, after applying the RWA, i.e., $H_{c,\text{cap}}
=g_{c,{\rm cap}}(\hat a_{c1}\hat a_{c2}^{\dag}+h.c.)$, and adding this term to the full
system Hamiltonian in Eq.~(\ref{eqA14}), we can find that the effective coupling strength can still
be expressed by Eq.~(\ref{eqA20}) with the $g_{c}=g_{c,{\rm cap}}-g_{c,{\rm ind}}$. This
result suggests that even accounting for the intermode capacitive coupling, the interaction
nulling condition is still independent on the fluxonium parameters. However, this cannot
explain the numerical result in the main text and the result shown in Fig.~\ref{setup_V2}(c), wherein the numeric study shows clear dependence of
the nulling condition on the fluxonium's transition frequency.

Alternatively, following the derivation without the RWA given in Sec.~\ref{IB2}, i.e., adding the capacitive
coupling term Eq.~(\ref{eqB1}) to the full system Hamiltonian in Eq.~(\ref{eqA21}), the previous study
of Ref.~\cite{Campbell2023} give the effective coupling strength of (keeping terms to
order $g_{c}/\Delta_{kl,i}$)
\begin{equation}
\begin{aligned}\label{eqB2}
&g_{kl,{\rm eff}}^{(D)}\approx\frac{g_{kl,1}g_{kl,2}}{2} \sum_{i=1,2} \frac{g_{c}(\Phi_{{\rm ext},c},\omega_{kl,i})}{\Delta_{kl,i}^{2}},\\
&g_{c}(\Phi_{{\rm ext},c},\omega_{kl,i})=(g_{c,{\rm cap}}-g_{c,{\rm ind}})-\frac{\Delta_{kl,i}}{\omega_{c}}(g_{c,{\rm cap}}+g_{c,{\rm ind}}).
\end{aligned}
\end{equation}
In contrast to Eq.~(\ref{eqA25}), this result shows that with the coexist of the intermode inductive and capacitive
couplings, the interaction nulling condition becomes frequency-dependent, rendering distinct biases to null
different fluxonium interactions and causing residual couplings, as illustrated in
Figs.~\ref{fig3}3(a) and~\ref{fig4}(a) of the main text for the Type-1 setup and in Fig.~\ref{setup_V2}(c) and~\ref{setup_V2}(f)
for the Type-2 setup.

To give further explicit illustrations, Figure~\ref{fig6} shows the residual couplings for the Type-2
setup (which exhibits more pronounced residuals, see Fig.~\ref{setup_V2}(f)) as a function of the qubit
bias with different strengths for the intermode capacitive coupling. Other system parameters are
the same as that used in Fig.~\ref{setup_V2}. By deceasing the strength of the intermode
couplings from 100 MHz to 0 MHz, the residuals also accordingly decreases to the level below 10 kHZ.
Furthermore, we also show the ZZ couplings, which remain consistently below 1 kHz, demonstrating that
the intermode capacitive coupling neither affect the fluxonium interactions within computational subspace
nor compromises the ZZ suppression.

\begin{figure}[htbp]
\begin{center}
\includegraphics[keepaspectratio=true,width=\columnwidth]{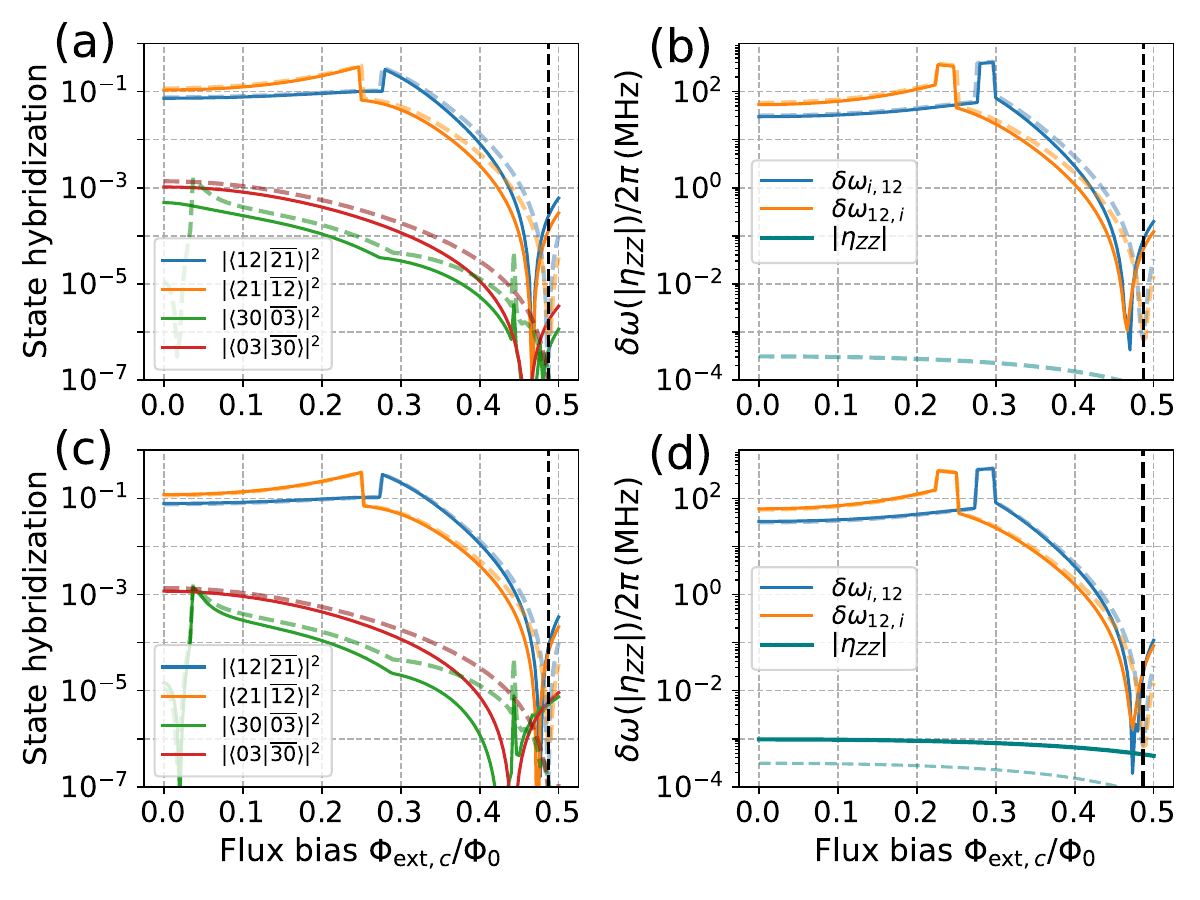}
\end{center}
\caption{The effect of the stray capacitive couplings on the coupler's functionality for
the Type-1 setup. (a,b) show the state hybridization and the state-dependent plasmon
frequency shifts versus the coupler flux bias with the intermode capacitive coupling strength
of $100\,\rm MHz$ and the NNN coupling strength of $25\,\rm MHz$. (c,d) is the same as in (a,b) but with
the intermode capacitive coupling strength of $100\,\rm MHz$ and the NNNN coupling
strength of $10\,\rm MHz$. (b) and (d) also display the ZZ coupling strengths. For comparative
analysis, the dashed curves show the results excluding NNN and NNNN couplings, see also
in Fig.~\ref{fig3} of the main text (vertical dashed lines indicate the coupler idle points that minimize the
frequency shifts). We note that sudden jumps or discontinuities in curves
are caused by state labeling failure near avoided crossings.}
\label{fig7}
\end{figure}

\begin{figure}[htbp]
\begin{center}
\includegraphics[keepaspectratio=true,width=\columnwidth]{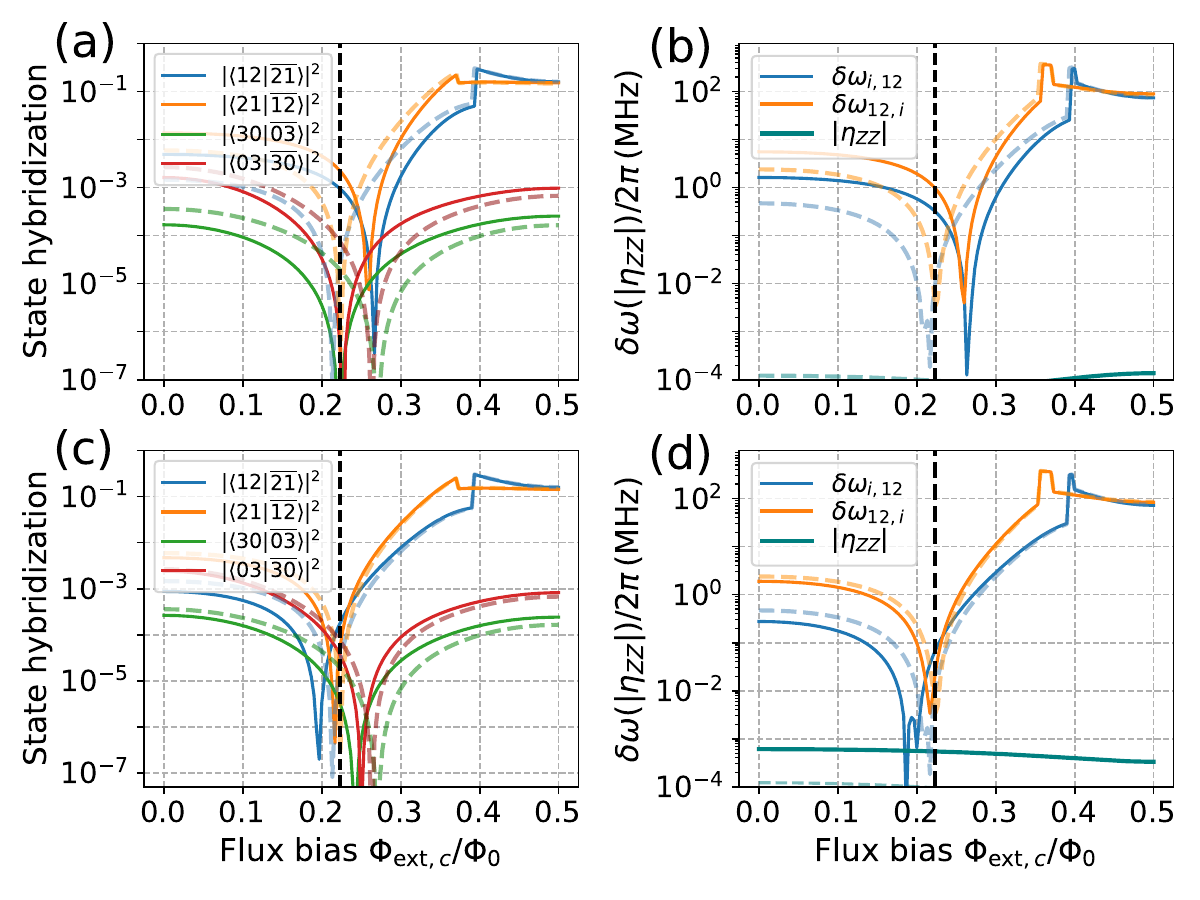}
\end{center}
\caption{The effect of the stray capacitive couplings on the coupler's functionality for
the Type-2 setup. We note that sudden jumps or discontinuities in curves
are caused by state labeling failure near avoided crossings.}
\label{fig8}
\end{figure}

\subsubsection{Stray capacitive couplings beyond nearest neighbors}\label{IC2}

In the current design, interactions beyond nearest neighbors can also exist via
stray capacitances or the indirect capacitive couplings mediated via the coupling circuit itself. For
the former one, the magnitude of the capacitance strongly depends on the specific geometry
layout of the qubit device, and can in principle be suppressed to a negligible level.
However, for the latter one, the magnitude can be non-negligible, especially for circuit
elements are coupled strongly with large coupling capacitors, as in the current design,
see Table~I of the main text.

For the coupling circuit shown in Fig.~\ref{fig1}(b) of the main text, the leading stray couplings beyond nearest
neighbors are the next-nearest-neighbor (NNN) coupling between $Q_{1}$ $(Q_{2})$ and
coupler mode $c2$ ($c1$), and the next-next-nearest-neighbor (NNNN) couplings for the two
fluxoniums, leading to the following coupling Hamiltonians,
\begin{equation}
\begin{aligned}\label{eqB3}
&H_{c,\text{NNN}}=J_{C1,\text{NNN}}\hat n_{1} \hat n_{c2}+J_{C2,\text{NNN}} \hat n_{2} \hat n_{c1},\\
&H_{c,\text{NNNN}}=J_{C,\text{NNNN}} \hat n_{1} \hat n_{2},
\end{aligned}
\end{equation}
where $J_{Ci,\text{NNN}}$ and $J_{C,\text{NNNN}}$ are the strengths of the NNN and NNNN couplings.
The magnitudes of the two-type couplings can be approximated by~\cite{Campbell2023,Yan2018,Galiautdinov2012}
\begin{equation}
\begin{aligned}\label{eqB4}
&J_{Ci,\text{NNN}}\approx \frac{2J_{ci}J_{C,12}}{\omega_{c}},\\
&J_{C,\text{NNNN}}\approx \frac{4J_{c1}J_{C,12}J_{c2}}{\omega_{c}^{2}}.
\end{aligned}
\end{equation}

Based on the circuit parameters in Table~\ref{tab:parameters} of the main text, we assume next-nearest-neighbor (NNN) and
next-next-nearest-neighbor (NNNN) coupling strengths of $J_{Ci,\text{NNN}}/2\pi=25\,{\rm MHz}$
and $J_{Ci,\text{NNN}}/2\pi=10\,{\rm MHz}$, respectively. Accordingly, accounting for the NNN and
the NNN couplings, Figures~\ref{fig7} and~\ref{fig8} show the state hybridization and the state-dependent
plasmon frequency shifts as functions of the coupler flux bias for the Type-1 and Type-2 setup, respectively.
For comparative analysis, the dashed curves show the results excluding NNN and NNNN couplings (vertical
dashed lines indicate the coupler idle points that minimize the
frequency shifts). While these stray NNN and NNNN couplings shift the interaction nulling points, they do not
alter the primary functions of the proposed coupler (i.e., programmable plasmon interactions with high on-off ratio)
and the main features without these NNN and NNNN stray couplings, see the dashed curves in Figs.~\ref{fig7} and~\ref{fig8}.
Additionally, ZZ couplings are also presented for the two setups, and as expected, the NNNN couplings (contributing
to direct fluxonium interactions) impact ZZ couplings more significantly than NNN couplings but do not
seriously affect the ZZ suppression (generally, the ZZ strength remains below 1 kHz).

\subsection{Variations in Josephson Junctions}\label{ID}

In the main text, the analysis assumes identical Josephson energies for the DTC's two transmons and
the dc-SQUID junctions. However, practical implementations face fabrication-induced parameter
variations.  Here, we turn to study how junction variations, including in transmon modes, SQUID junctions,
and fluxonium's main junction, affect the coupling circuit. In the following discussion, we consider
that given the proven available state-of-the-art, the parameter accuracy of josephson junctions is well
within $10\%$~\cite{Kreikebaum2020}.

\begin{figure}[htbp]
\begin{center}
\includegraphics[keepaspectratio=true,width=\columnwidth]{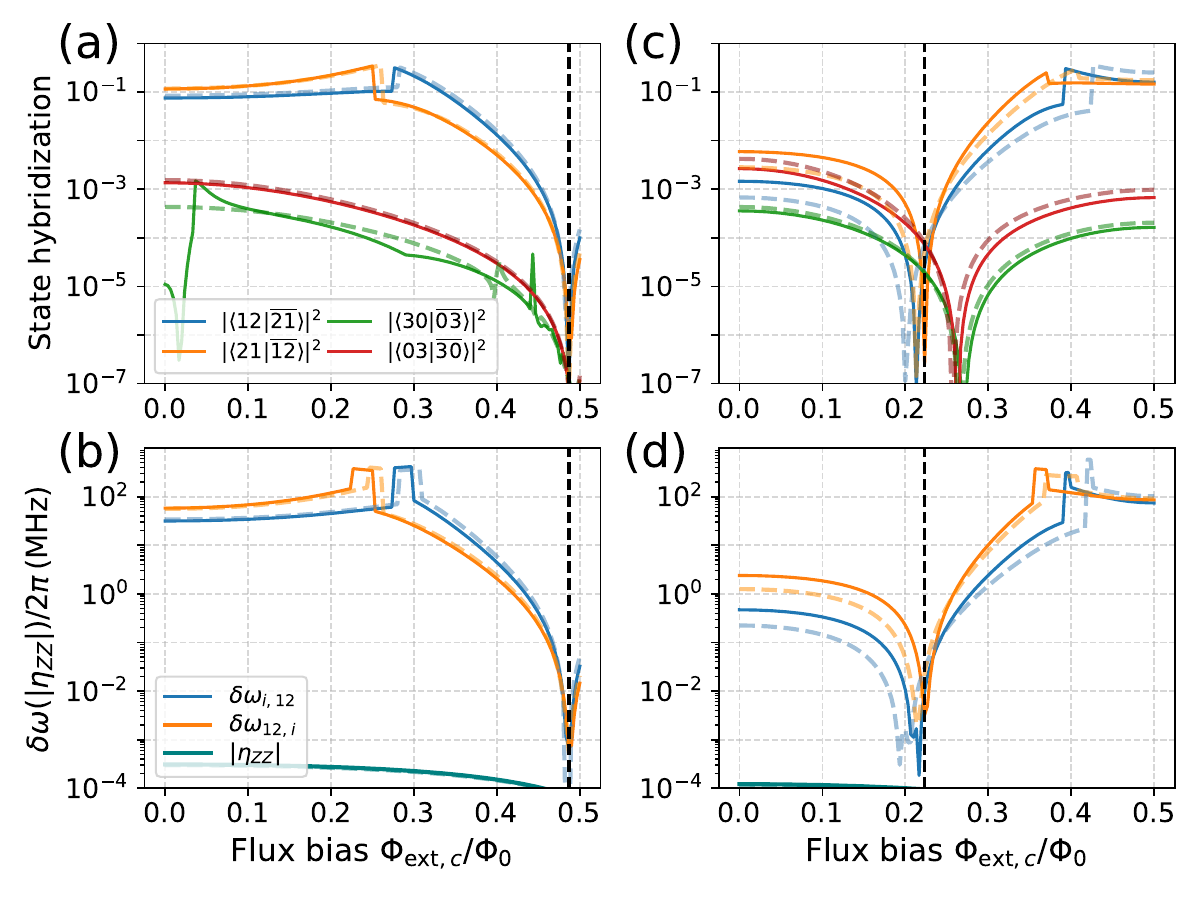}
\end{center}
\caption{The effect of junction variations in two transmon modes on the coupler. (a,b) and (c,d) show the state
hybridization and the state-dependent plasmon frequency shifts versus the coupler flux bias with
$(E_{J,c2}-E_{J,c1})/E_{J,c1}=10\%$ for the Type-1 and Type-2
setup, respectively. (b) and (d) also display the ZZ coupling strengths. For comparison, the dashed
curves show results without the junction variation. We note that sudden jumps or discontinuities in curves
are caused by state labeling failure near avoided crossings.}
\label{fig9}
\end{figure}
\subsubsection{Junction variations in two transmon modes}\label{ID1}

Considering the junction variations, Figure~\ref{fig9} shows both the state
hybridization and the state-dependent plasmon frequency shifts versus the coupler flux
bias with non-degenerate coupler modes, i.e., $(E_{J,c2}-E_{J,c1})/E_{J,c1}=10\%$.
Other parameters are the same as in the main text. For easy reference, results excluding the variation
are also presented, see the dashed curves in Fig.~\ref{fig9}. Notably, both the Type-1 (see Fig.~\ref{fig9}(a)) and Type-2
setup (see Fig.~\ref{fig9}(b)) maintain their main functionality (i.e., programmable plasmon interactions
with high on-off ratio) despite these variations, demonstrating the robustness of the coupler design.

\begin{figure}[htbp]
\begin{center}
\includegraphics[keepaspectratio=true,width=\columnwidth]{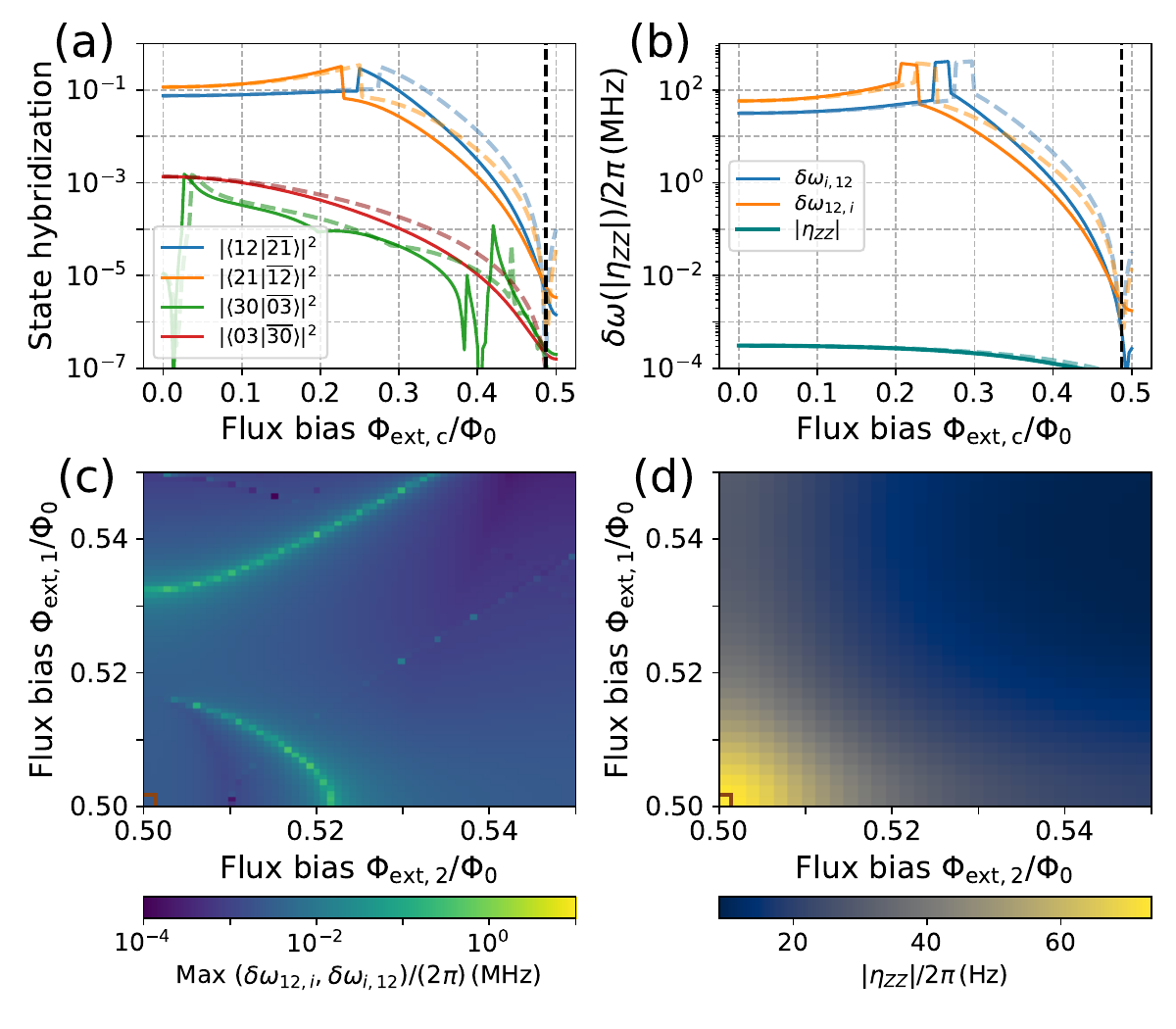}
\end{center}
\caption{The effect of junction asymmetry of the SQUID on the Type-1 setup with the consideration of
the intrinsic flux crosstalk between the main and the flux loops. Here, the junction symmetry is assumed
to be $(E_{J,ca}-E_{J,cb})/E_{J,cb}=10\%$ (see Fig.~\ref{fig5}). (a) and (b) show the state
hybridization and the state-dependent plasmon frequency shifts versus the coupler flux
bias, respectively. For easy reference, dashed curves show the results excluding the junction
asymmetry and the flux crosstalk. (c) and (d) display the residual couplings in the non-computational and
computational subspace quantified by the state-dependent frequency shift and the ZZ couplings. We note
that in (a,b), sudden jumps or discontinuities in curves are caused by state labeling failure
near avoided crossings.}
\label{fig10}
\end{figure}

\subsubsection{Junction asymmetry in the dc SQUID}\label{ID2}

According to the discussion given in Sec.~\ref{IA}, for the Type-1 setup, the junction asymmetry in the SQUID
can not only compromise the tunable range of the intermode inductive couplings, see Eq.~(\ref{eqA3}), but
also can affect the intrinsic flux crosstalk between the main and SQUID loops, see Eq.~(\ref{eqA4}). In
the main text, the analysis assume that the junctions are identical and the intrinsic crosstalk is
compensated for allowing independent flux bias (by actively eliminating the crosstalk or
passively suppressing it via the gradiometric design~\cite{Paauw2009}). By relaxing the two
assumptions, Figures~\ref{fig10}(a) and~\ref{fig10}(b) show the state hybridization and the state-dependent plasmon
frequency shifts versus the coupler flux bias with the consideration of both the junction asymmetry
$(E_{J,ca}-E_{J,cb})/E_{J,cb}=10\%$ (see Fig.~\ref{fig5}) and the intrinsic flux crosstalk. For comparison, results excluding
the junction asymmetry and the flux crosstalk are also presented, see the dashed curves and also shown
in the main text. Additionally, Figures~\ref{fig10}(c) and~\ref{fig10}(d) also display
the residual couplings as function of the qubit flux bias. All these results illustrate that even
considering the junction asymmetry and the intrinsic flux crosstalk, the Type-1 setup can still preserve its main
functionality (i.e., programmable plasmon interactions with high on-off ratio), demonstrating the robustness
of this operational design to the junction asymmetry and the intrinsic crosstalk.

\begin{figure}[htbp]
\begin{center}
\includegraphics[keepaspectratio=true,width=\columnwidth]{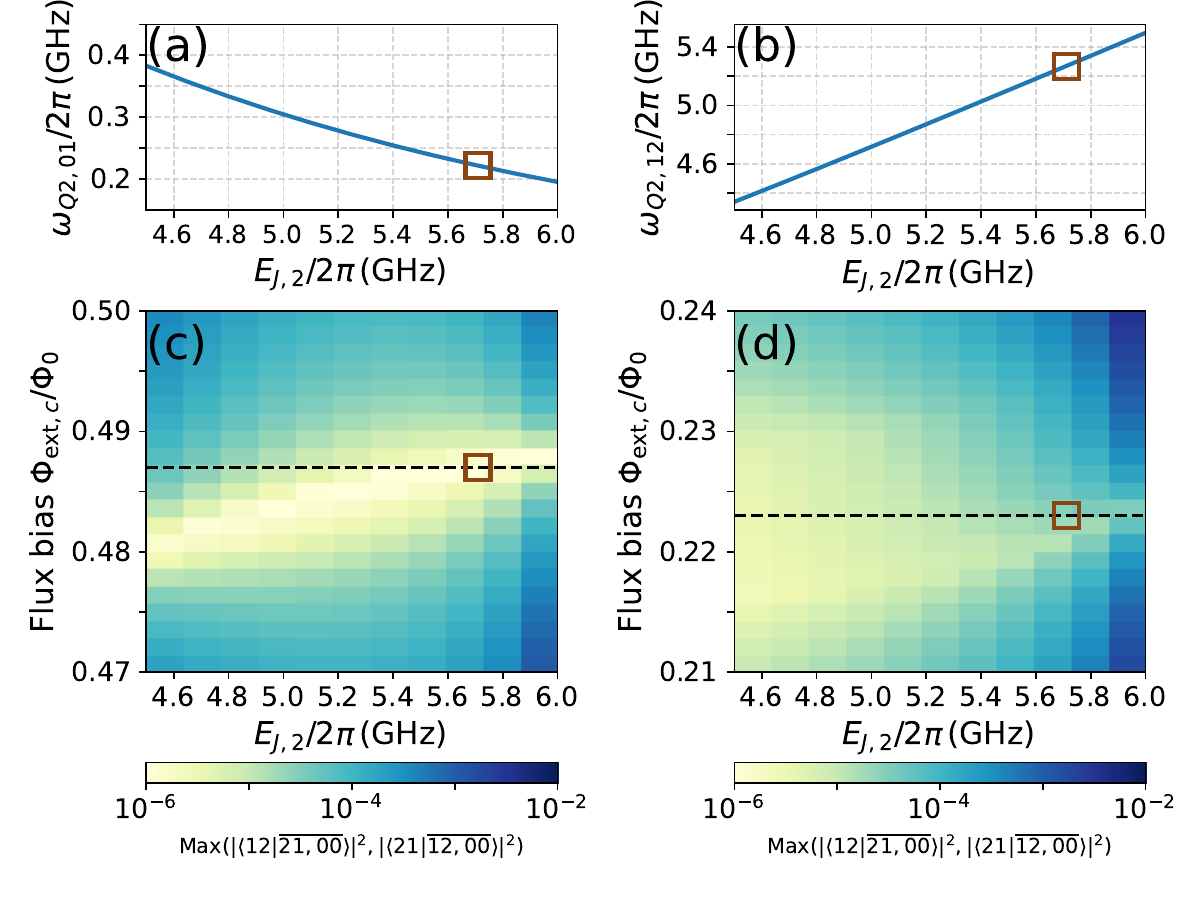}
\end{center}
\caption{The sensitivity of interaction nulling points on the fluxonium's Josephson energies.
(a) and (b) show the fluxonium frequency, i.e., $\omega_{01}$ and $\omega_{12}$, as function of
$Q_{2}$'s Josephson energies $E_{J,2}$. (c) and (d) present the state hybridization, ${\rm Max}(|12\langle|\overline{21,00}\rangle|^2,\,|21\langle|\overline{12,00}\rangle|^2)$, as
functions of the coupler flux bias and $E_{J,2}$ for the Type-1 and Type-2
setup, respectively. Other used parameters are same as in Table~\ref{tab:parameters} of
the main text. In (a,b), open squares indicate the fluxonium parameters studied in the main text
and in (c,d), open squares and horizonal dashed lines marks the corresponded interaction
nulling point.}
\label{fig11}
\end{figure}

\subsubsection{Varying Josephson energies of the fluxonium}\label{ID3}

As demonstrated both in the main text and in Sec.~\ref{IIA}, the inclusion of intermode capacitive coupling
renders the plasmon interaction nulling condition frequency-dependent. We now turn to investigate the sensitivity
of this nulling point to variations in the fluxonium's Josephson energies. Figures~\ref{fig11}(a) and~\ref{fig11}(b)
display the corresponding fluxonium frequency dependence on $Q_{2}$ Josephson energies $E_{J,2}$, with other parameters
same as in Table~I of the main text. Accordingly, Figures~\ref{fig11}(c) and~\ref{fig11}(d) show
the state hybridization, ${\rm Max}(|12\langle|\overline{21,00}\rangle|^2,\,|21\langle|\overline{12,00}\rangle|^2)$, as
functions of the coupler flux bias and $E_{J,2}$ for both the two operational setups. For each $E_{J,2}$, the
interaction nulling point corresponds to the location where the state hybridization is minimized.
Generally, these results demonstrate that the interaction nulling condition only shows a weak dependence
on the variation of fluxonium's Josephson energies.

We note that here the fluxoniums are still
biased at their half-flux sweet spots in contrast to the case studied in Fig.~4 of the main text, where the fluxoniums are
biased away from the sweet spot, breaking down the selection rule at half-integer flux sweet spots.

\subsection{Residual Couplings and frequency collisions}\label{IE}

As shown in Fig.~3 of the main text and in Fig.~\ref{setup_V2}, when accounting for intermode capacitive coupling, the
interaction nulling condition in the Type-1 setup exhibits significantly weaker frequency dependence
compared to that in the Type-2 setup. Furthermore, as shown in Fig.~\ref{setup_V2}(f), the type-2 setup generates
more substantial residual couplings when fluxoniums are biased away from their half-flux sweet spots.
Here, we show that these operational contrasts can be largely attributed to frequency collisions involving
coupler modes. Thus, to suppress the residuals, we need to operate the fluxonium's plasmon transitions
detuned from coupler modes.

\begin{figure}[htbp]
\begin{center}
\includegraphics[keepaspectratio=true,width=\columnwidth]{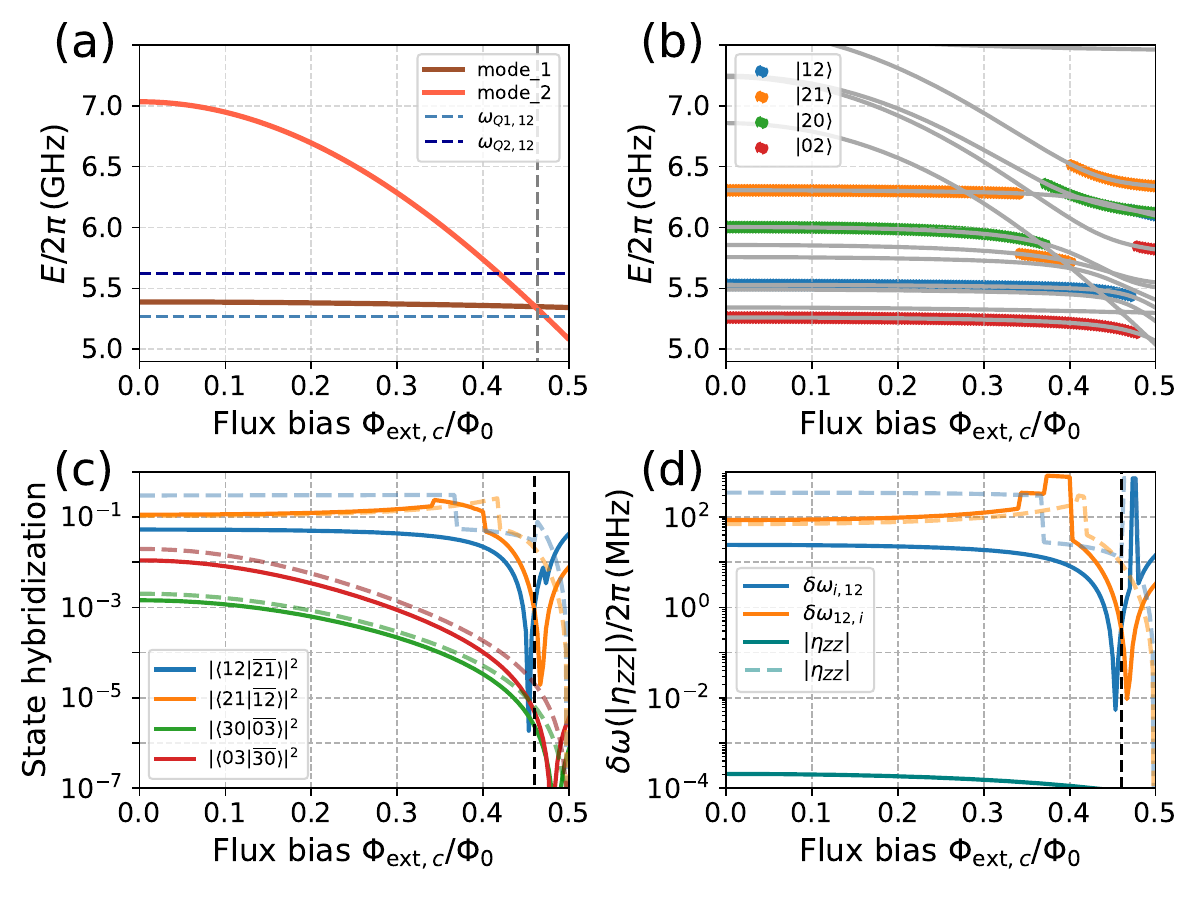}
\end{center}
\caption{Coupler eigenmode frequencies (a), energy levels of the coupled fluxonium circuit (b), the
state hybridization (c), and the state-dependent plasmon frequency shifts (d) as functions of the
coupler flux bias for the Type-1 setup with the transmon's Josephson energy of $15\,\rm GHz$. Other parameters
are same as in the main text, see Table~\ref{tab:parameters} of the main text. In (a), the grey vertical dashed line
indicates the intermode interaction nulling point for coupling circuit excluding the parts of the
fluxoniums. In (c) and (d), for comparison, dashed lines indicate the results excluding the intermode
capacitive couplings, and the black vertical dashed line mark the intermode interaction nulling
point for the full system (i.e., minimizing the state dependent shifts). We note that in (c,d), sudden
jumps or discontinuities in curves are caused by state labeling failure near avoided crossings.}
\label{fig16}
\end{figure}

\subsubsection{The Type-1 setup}\label{IE1}

As the same as in Figs.~\ref{fig2}(a,b) and~\ref{fig3}(a,b) of the main text, Figure~\ref{fig16} shows
results, including coupler eigenmode frequencies, energy levels of the coupled fluxonium circuit, the
state hybridization, and the state-dependent plasmon frequency shifts as functions of the
coupler flux bias, for the Type-1 setup with the transmon's Josephson energy of $15\,\rm GHz$. Other parameters
are same as that in the main text, see Table~\ref{tab:parameters} of the main text. As shown in
Fig.~\ref{fig16}(a), increasing the transmon's Josephson energy from 9 GHz to 15 GHz brings one coupler
eigenmode near resonance with both fluxoniums' plasmon transitions ($|1\rangle\leftrightarrow|2\rangle$).
Consequently, in contrast to the case studied in the main text (see Fig.~\ref{fig3}(a) of the main text), the
interaction nulling conditions show more significant frequency dependence on the plasmon transition
frequency, see Fig.~\ref{fig16}(c). For comparison, dashed lines in Figs.~\ref{fig16}(c) and~\ref{fig16}(d)
indicate the results excluding the intermode capacitive couplings.

\begin{figure}[htbp]
\begin{center}
\includegraphics[keepaspectratio=true,width=\columnwidth]{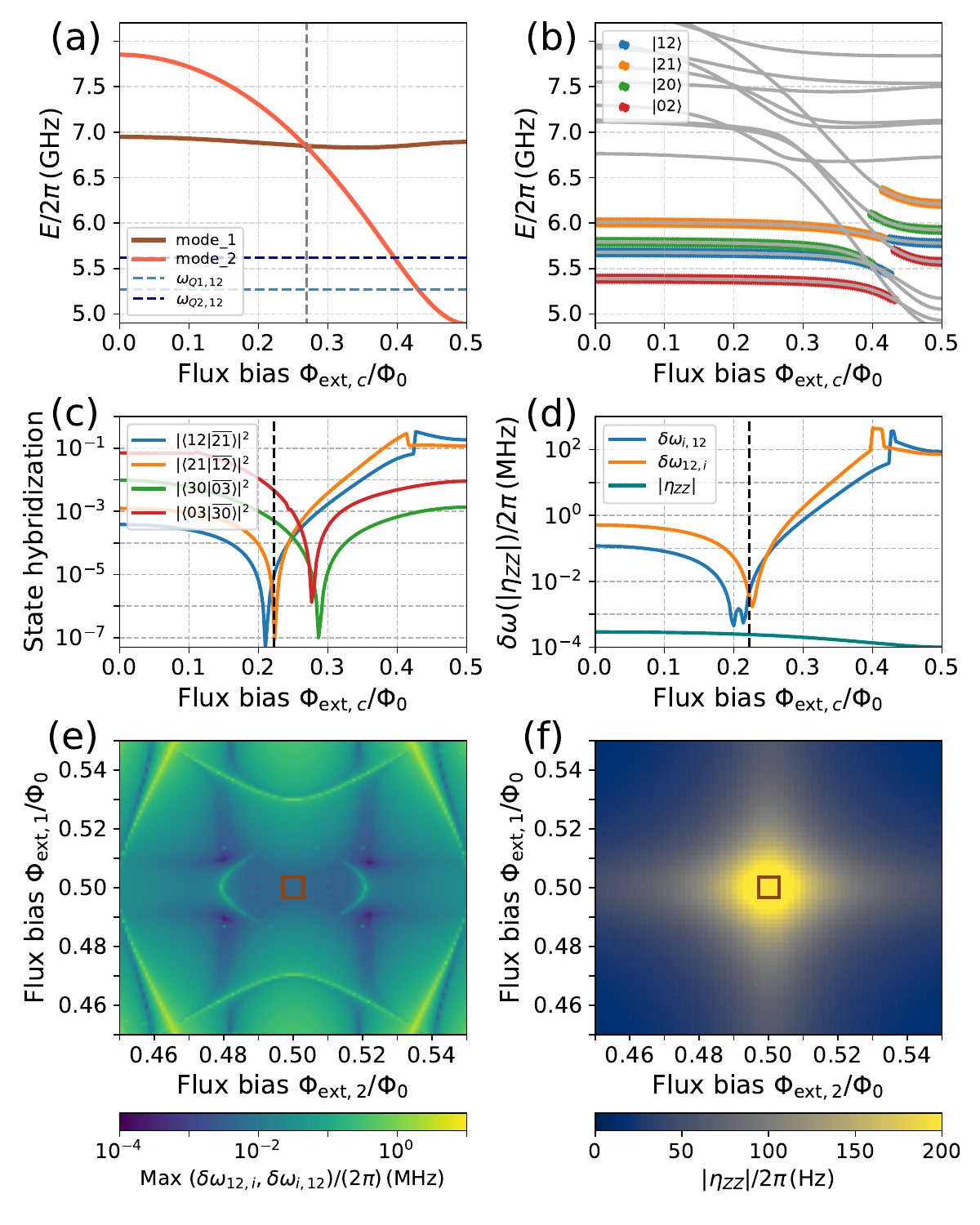}
\end{center}
\caption{Tunable plasmon interactions and the residuals in the Type-2 setup with new coupler
parameters, i.e., the Set-A of Table~\ref{tab:parameters2}. Other circuit parameters are the
same as that used in Fig.~\ref{setup_V2}. We note that in (c,d), sudden jumps or discontinuities in curves
are caused by state labeling failure near avoided crossings.}
\label{fig17}
\end{figure}

\begin{figure}[htbp]
\begin{center}
\includegraphics[keepaspectratio=true,width=\columnwidth]{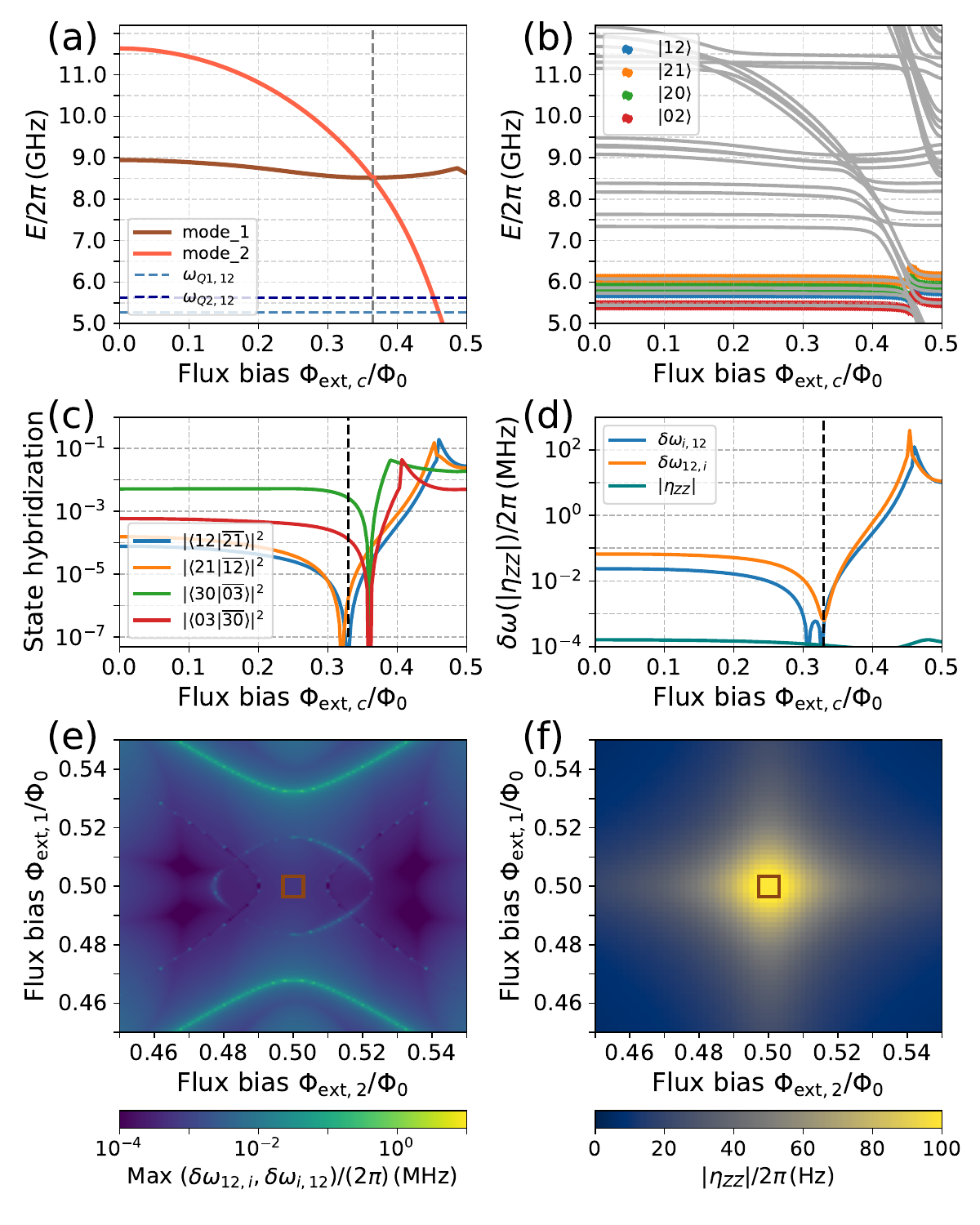}
\end{center}
\caption{Tunable plasmon interactions and the residuals in the Type-2 setup with new coupler
parameters, i.e., the Set-B of Table~\ref{tab:parameters2} . Other circuit parameters are the
same as that used in Fig.~\ref{setup_V2}. We note that in (c,d), sudden jumps or discontinuities in curves
are caused by state labeling failure near avoided crossings.}
\label{fig18}
\end{figure}

\subsubsection{The Type-2 setup}\label{IE2}

As mentioned before, to suppress the residuals, the coupler modes should be pushed away from the
plasmon transitions of fluxonium qubits. For subsequent analysis, we maintain fixed fluxonium parameters
while optimizing coupler design parameters to minimize residual interactions.

For the Type-2 setup, Figures~\ref{fig17} and~\ref{fig18} show
the results with two new parameter sets of the coupler circuit, as summarized in
Table~\ref{tab:parameters2}. Other circuit parameters are the same as in Table~\ref{tab:parameters}
of the main text. Similar to the residual suppression by reducing intermode
capacitive coupling (see Fig.~\ref{fig6}), by increasing the coupler-plasmon detuning, the residuals also
accordingly decreases to the level below 10kHZ, see Figs.~\ref{fig17}(e)
and~\ref{fig18}(e). Furthermore, we also show the ZZ couplings, which remain
below 1 kHz, as shown in Figs.~\ref{fig17}(f) and~\ref{fig18}(f).

\begin{table}[!htb]
\caption{\label{tab:parameters2} Two parameter sets (denoted as set-A and set-B) for the coupler circuit, which is used
for the results shown in Figs.~\ref{fig17} and~\ref{fig18}. }
\begin{ruledtabular}
\begin{tabular}{cccc}
$ $&
$E_C$ (GHz)&
$E_J$ (GHz)&
$E_{J,12}$ (GHz)\\\hline
Set-A & 0.20 & 30  & 7  \\\hline
Set-B & 0.25 & 40 & 18 \\
\end{tabular}
\end{ruledtabular}
\end{table}

\subsection{Coupler-induced qubit decoherence}\label{IF}

For any proposed coupling architecture, two competing requirements must be balanced, i.e., sufficiently strong
coupling to enable fast gate operations while minimizing coupling-induced relaxation and dephasing on the
qubit. Here, we turn to study the coupler-induced relaxation and dephasing on both the computational states
and non-computational gate states in the proposed coupling architecture.

\begin{figure}[htbp]
\begin{center}
\includegraphics[keepaspectratio=true,width=\columnwidth]{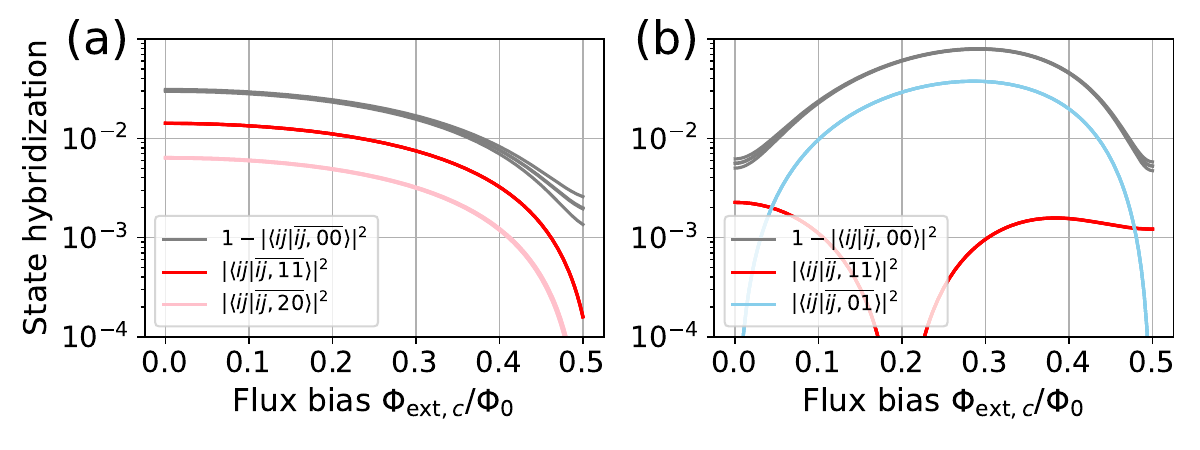}
\end{center}
\caption{The dressing of the computational states from the fluxonium-coupler couplings,
quantified by the overlap between the computational states and the associated bare state. (a)
and (b) are for the Type-1 and Type-2 setup, respectively. Here, the label $ij$ runs
over all the computational states, i.e., $(00,\,01,\,10,\,11)$. }
\label{fig12}
\end{figure}

\begin{figure}[htbp]
\begin{center}
\includegraphics[keepaspectratio=true,width=\columnwidth]{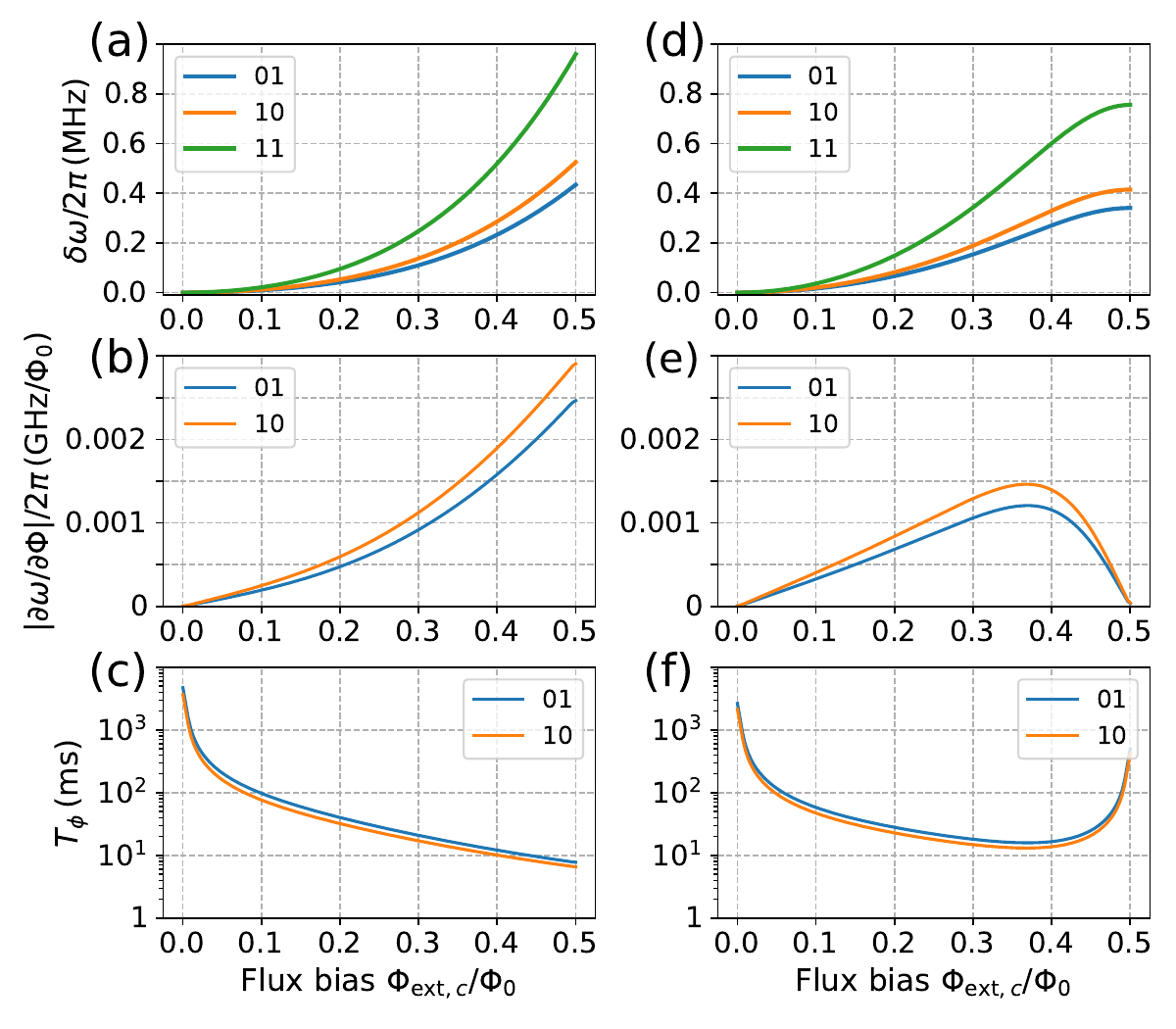}
\end{center}
\caption{The coupler-induced frequency shift on the computational levels $\delta\omega$ (a,d), the sensitivity
of the energies of the computational levels to the coupler flux
bias $\partial\omega/\partial\Phi$ (b,e), and the coupler-induced pure dephasing time as functions of the
coupler flux bias (c,f). (a-c) and (d-f) are for the for the Type-1 and Type-2 setup, respectively.
Here, the dephasing times are estimated by assuming 1/f noise with the noise amplitude of $10\,\mu\Phi_{0}$.}
\label{fig13}
\end{figure}

\subsubsection{Decoherence in computational subspace}\label{IF1}

As demonstrated in Fig.~\ref{fig2}(b) of the main text and in Fig.~\ref{setup_V2}(b), the computational subspace exhibits
negligible interaction with the coupler, enabled by the weak dipole moment of the qubit transition
within computational subspace and the low transition frequencies. This decoupling can be
further verified by the dressing of the computational
subspace from the fluxonium-coupler couplings. Figures~\ref{fig12}(a) and~\ref{fig12}(b) show
the dressing of the qubit states, quantified by the overlap between the computational states
and the associated bare state, as function of the coupler flux bias for the Type-1
and Type-2 setup, respectively. The near-complete independence of dressing on the
the qubit states (manifested as coincident curves for any specific coupler state, see Fig.\ref{fig12}) confirms
that the qubit subspace is almost complete decoupled from the coupler, suppressing the coupler-induced
qubit Purcell decay to negligible levels.

Furthermore, Figure~\ref{fig13} shows the coupler-induced frequency shift on the computational
levels $\delta\omega$, the sensitivity of the energies of the computational levels to the coupler flux
bias $\partial\omega/\partial\Phi$, and the coupler-induced pure dephasing
time of $T_{\phi,{\rm 1/f}}=1/(\sqrt{A_{\Phi}^2\ln{2}}|\partial\omega/\partial\Phi|)$ (assuming $1/f$ noise
with the typical flux noise amplitude $A_{\Phi}$ of $10\,\mu\Phi_{0}$~\cite{Koch2007}) as functions of the coupler
flux bias for both two setups. These results show that the coupler-induced dephasing
time is generally above 10 $\rm ms$ across the entire bias range.

Overall, we conclude that due to the weak dipole moment of the qubit transition with the extremely
low transition frequencies, the proposed coupling architecture can preserve the demonstrated
state-of-the-art high coherence in the computational subspace.

\begin{figure}[htbp]
\begin{center}
\includegraphics[keepaspectratio=true,width=\columnwidth]{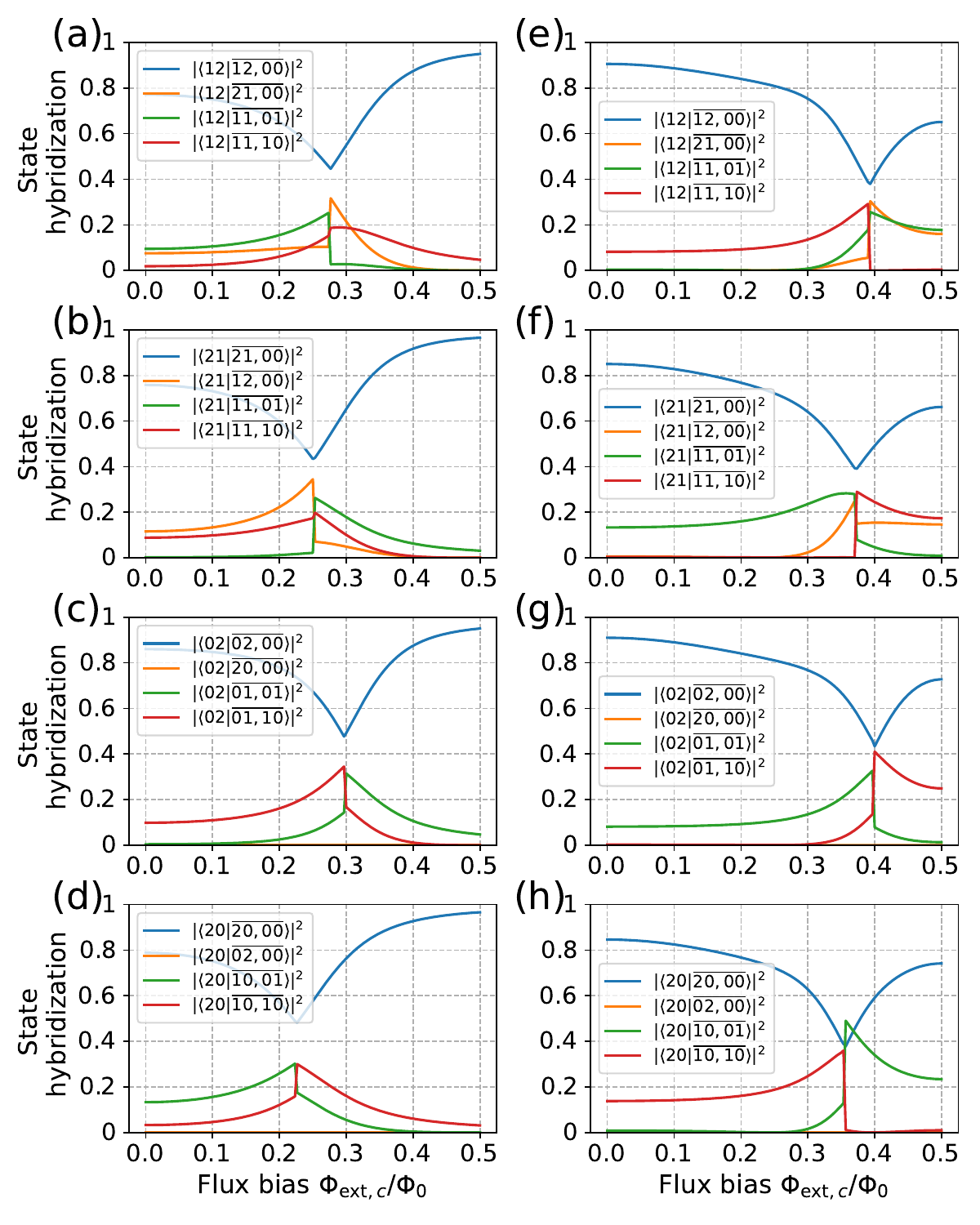}
\end{center}
\caption{The dressing of the non-computational gate states from the interactions between
the plasmon transitions ($|1\rangle\leftrightarrow|2\rangle$) and the coupler modes,
quantified by the overlap between the computational states and the associated bare state. (a-d)
and (e-h) are for the Type-1 and Type-2 setup, respectively. We note that sudden jumps or discontinuities in curves
are caused by state labeling failure near avoided crossings.}
\label{fig14}
\end{figure}

\begin{figure}[htbp]
\begin{center}
\includegraphics[keepaspectratio=true,width=\columnwidth]{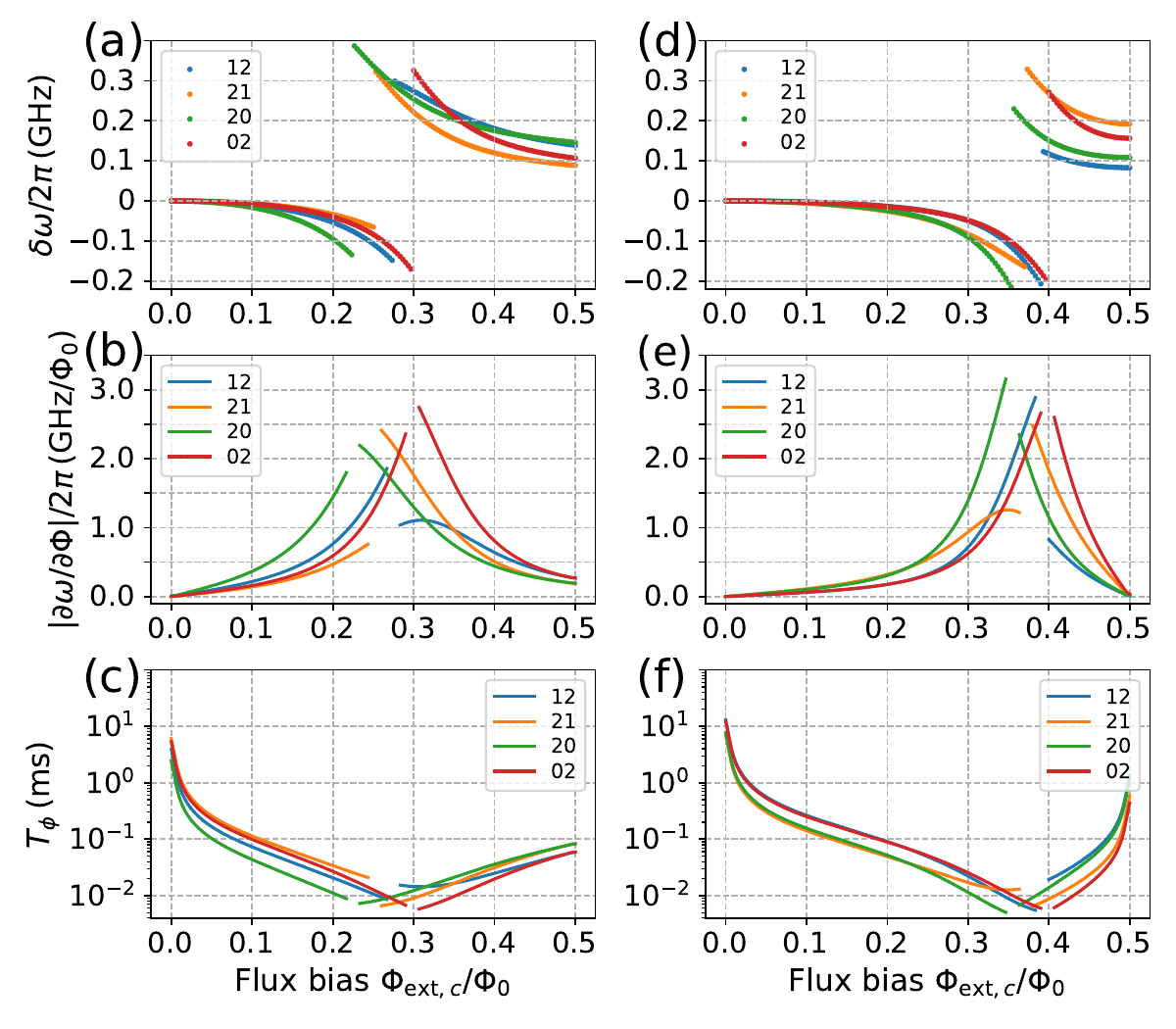}
\end{center}
\caption{The coupler-induced frequency shift $\delta\omega$ (a,d), the flux sensitivity
$\partial\omega/\partial\Phi$ (b,e), and the coupler-induced dephasing time for the
non-computational levels versus the coupler flux bias (c,f). (a-c) and (d-f) are for the for the Type-1 and Type-2 setup, respectively.
Here, the dephasing times are estimated by assuming 1/f noise with the noise amplitude of $10\,\mu\Phi_{0}$. We note
that discontinuities in curves
are caused by state labeling failure near avoided crossings.}
\label{fig15}
\end{figure}

\subsubsection{Decoherence for non-computational gate states}\label{IF2}

In the proposed architecture, temporarily occupying levels outside the
computational subspace offers the key to realize fast two-qubit gates. For gate
schemes involving temporary population of non-computational states, the main gate error
could be from the relaxation and dephasing of the non-computational states. Thus, here
we turn to investigate coupled-induced relaxation and dephasing on the non-computational
levels, i.e., $|02(20)\rangle$ and $|12(21)\rangle$.

Figure~\ref{fig14} shows the dressing of these non-computational states, quantified by
the overlap between the non-computational states and the associated bare state (due to the interactions
between the plasmon transition ($|1\rangle\leftrightarrow|2\rangle$) and the coupler modes), as function of
the coupler flux bias. Generally, the overlap is below $20\%$ for both two setups, leading to the Purcell-limited
relaxation time of $\sim 5T_{1,c}$ ($T_{1,c}$  denotes the life time of coupler modes). Considering that
the state-of-the-art transmon's relaxation time generally exceeds $100\,\rm \mu s$, we thus optimistically expect
that the coupler-induced Purcell decay should not limit the lifetime of these non-computational levels (currently, the
typical lifetime of these levels is about $10\,\rm \mu s$~\cite{Ficheux2021,Ding2023}).

Additionally, similar to Fig.~\ref{fig13}, Figure~\ref{fig15} shows the results including the frequency
shift $\delta\omega$, the flux sensitivity $\partial\omega/\partial\Phi$, and the coupler-induced dephasing
time for the non-computational gate states. By assuming $1/f$ noise with the typical noise
amplitude of $10\,\mu\Phi_{0}$, the coupler-induced dephasing time is generally
above $10\,\rm \mu s$ and when biasing the system away from the frequency collision associated with the
on-resonance plasmon-coupler interactions, the dephasing time can excess $100\,\rm \mu s$. As
discussed in Sec.~\ref{IIIE}, for 50ns-CZ gates, the decoherence gate error is
below $10^{-5}$ ($10^{-7}$) for the dephasing time of $10\,\rm \mu s$ ($100\,\rm \mu s$).

\section{The fluxonium architecture with single-transoms couplers}\label{II}

Here, we begin by giving the derivations of the full system Hamiltonian and the effective Hamiltonian
for the architecture based on the single-transmon couplers (STC), as shown in Fig.~\ref{fig1}(c) of the main text, and
then evaluate the coupler-induced qubit decoherence. Hereafter, to describe the system state within the STC
setup, we use the notation of $|Q_{1}Q_{2},C\rangle$ and when confined to qubit subspace,
notation $|Q_{1}Q_{2}\rangle$ is used for $|Q_{1}Q_{2},0\rangle$.

\subsection{Coupler-mediated tunable couplings}\label{IIA}

The fluxonium architecture based on the STC can be modeled by the following Hamiltonian
\begin{equation}
\begin{aligned}\label{eqE1}
H^{(S)}=& \sum_{i=1,2}[4 E_{C,i} \hat n^2_i + \frac{E_{L,i}}{2}(\hat\varphi_i - \varphi_{\text{ext},i})^2 - E_{J,i}\cos\hat\varphi_i]\\
&+J_{c1}\hat n_1 \hat n_{c}+J_{c2}\hat n_2 \hat n_{c}+J_{12}\hat n_1 \hat n_{2}
\\&+4 E_{C,c} \hat n^2_{c} - E_{J,c}\cos(\frac{\varphi_{\text{ext},c}}{2})\cos\hat\varphi_{c},
\end{aligned}
\end{equation}
Following the derivation in Sec.~\ref{IB}, i.e., approximating the STC as a harmonic mode and focusing on
one specific fluxonium's transition $|k\rangle\leftrightarrow|l\rangle$, the system Hamiltonian
can be approximated by
\begin{equation}
\begin{aligned}\label{eqE2}
H_{kl}^{(S)}=&\sum_{i=1,2}[\omega_{kl,i}\hat\sigma_{kl,i}^{\dag}\hat\sigma_{kl,i}+g_{kl,i}(\hat\sigma_{kl,i}^{\dag}\hat a_{c}+h.c.)]
\\&+\omega_{c}\hat a_{c}^{\dag}\hat a_{c}+g_{kl,12}(\hat\sigma_{kl,1}^{\dag}\hat\sigma_{kl,2}+h.c.).
\end{aligned}
\end{equation}

Again, by assuming that the system operates in the dispersive regime, i.e., the interaction
strength $g_{kl,i}$ significantly smaller than the fluxonium-coupler
detuning $\Delta_{kl,i}=\omega_{kl,i}-\omega_{c}$, an effective Hamiltonian can be obtained
by eliminating the direct fluxonium-coupler interactions, giving rise to

\begin{equation}
\begin{aligned}\label{eqE3}
&H_{kl}^{({\rm eff})}= \sum_{i=1,2}[\omega_{kl,i}\hat\sigma_{kl,i}^{\dag}\hat\sigma_{kl,i}]+g_{kl,{\rm eff}}^{(S)}(\hat\sigma_{kl,1}^{\dag}\hat\sigma_{kl,2}+h.c.),
\\&g_{kl,{\rm eff}}^{(S)}\approx g_{kl,12}+\frac{g_{kl,1}g_{kl,2}}{2}\left[\frac{1}{\Delta_{kl,1}}+\frac{1}{\Delta_{kl,2}}\right],
\end{aligned}
\end{equation}

As suggested by Eq.~(\ref{eqE3}), the coupler-mediated interaction in the STC setup exhibits significantly stronger
dependence on fluxonium transition frequencies compared to the DTC case, see Eq.~(\ref{eqA25}). Thus, the STC
setup generally requires distinct bias points to null different plasmon transitions. Given that interaction
sensitivity at nulling points scales as $\sim 1/\Delta_{kl,i}^{2}$, the frequency dependence on fluxonium transitions
can be minimized by increasing both the direct fluxonium-fluxonium coupling $J_{12}$ and the fluxonium-coupler
detuning, thereby simultaneously suppressing all dominant inter-fluxonium interactions to experimentally achievable
levels, as demonstrated in the main text, see Fig.~\ref{fig3}(e).

\subsection{Coupler-induced qubit decoherence}\label{IIB}

\begin{figure}[htbp]
\begin{center}
\includegraphics[keepaspectratio=true,width=\columnwidth]{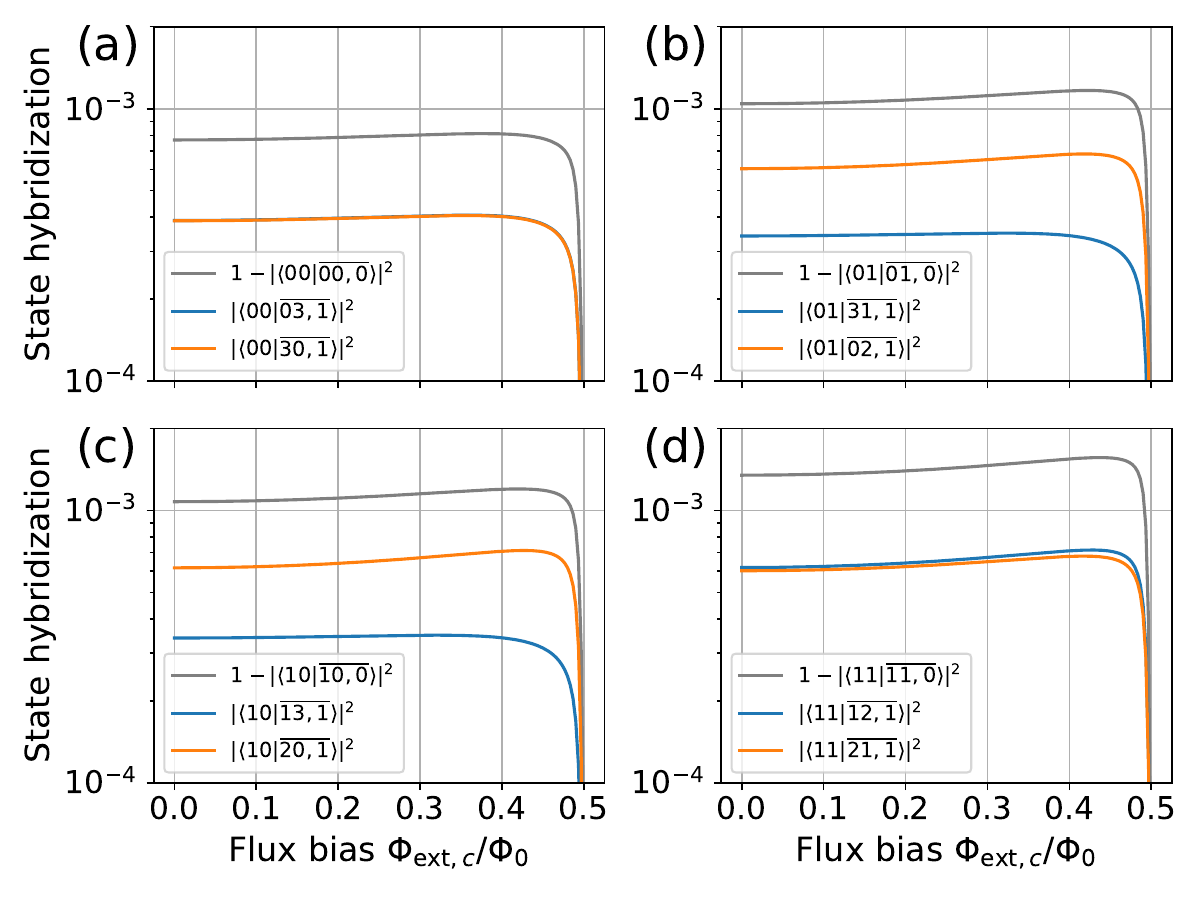}
\end{center}
\caption{The dressing of the computational states, quantified by the overlap between the computational states
and the associated bare state, in the architecture based on the STC. (a-d) are for the four computational
states $(00,\,01,\,10,\,11)$. }
\label{STC_qubit_overlap}
\end{figure}

\begin{figure}[htbp]
\begin{center}
\includegraphics[keepaspectratio=true,width=\columnwidth]{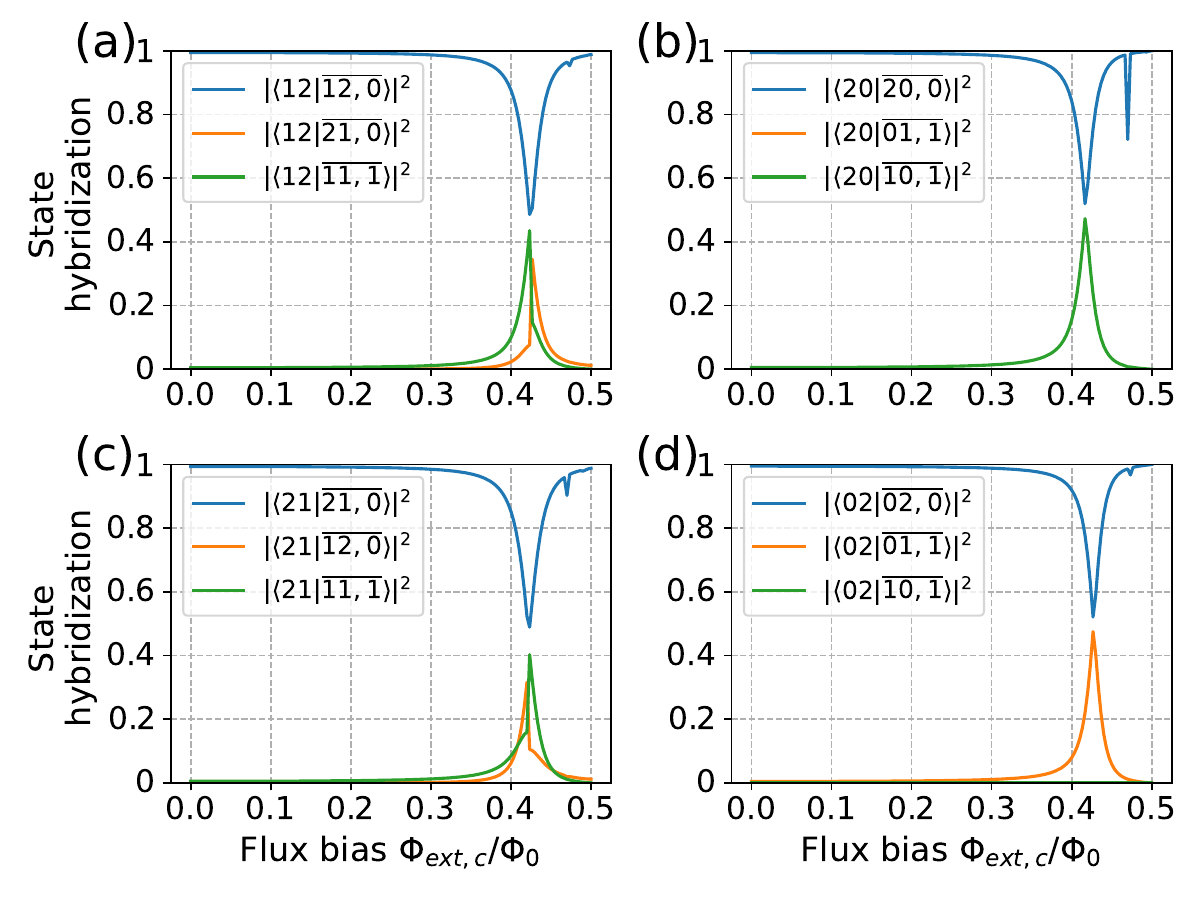}
\end{center}
\caption{The dressing of the non-computational gate states from the interactions between
the plasmon transitions ($|1\rangle\leftrightarrow|2\rangle$) and the coupler modes,
quantified by the overlap between the computational states and the associated bare state
in the architecture based on the STC. (a-d) are for the four possible non-computational
gate states $(12,\,20,\,21,\,02)$. We note that sudden jumps or discontinuities in curves
are caused by state labeling failure near avoided crossings.}
\label{STC_gate_overlap}
\end{figure}

\begin{figure}[htbp]
\begin{center}
\includegraphics[keepaspectratio=true,width=\columnwidth]{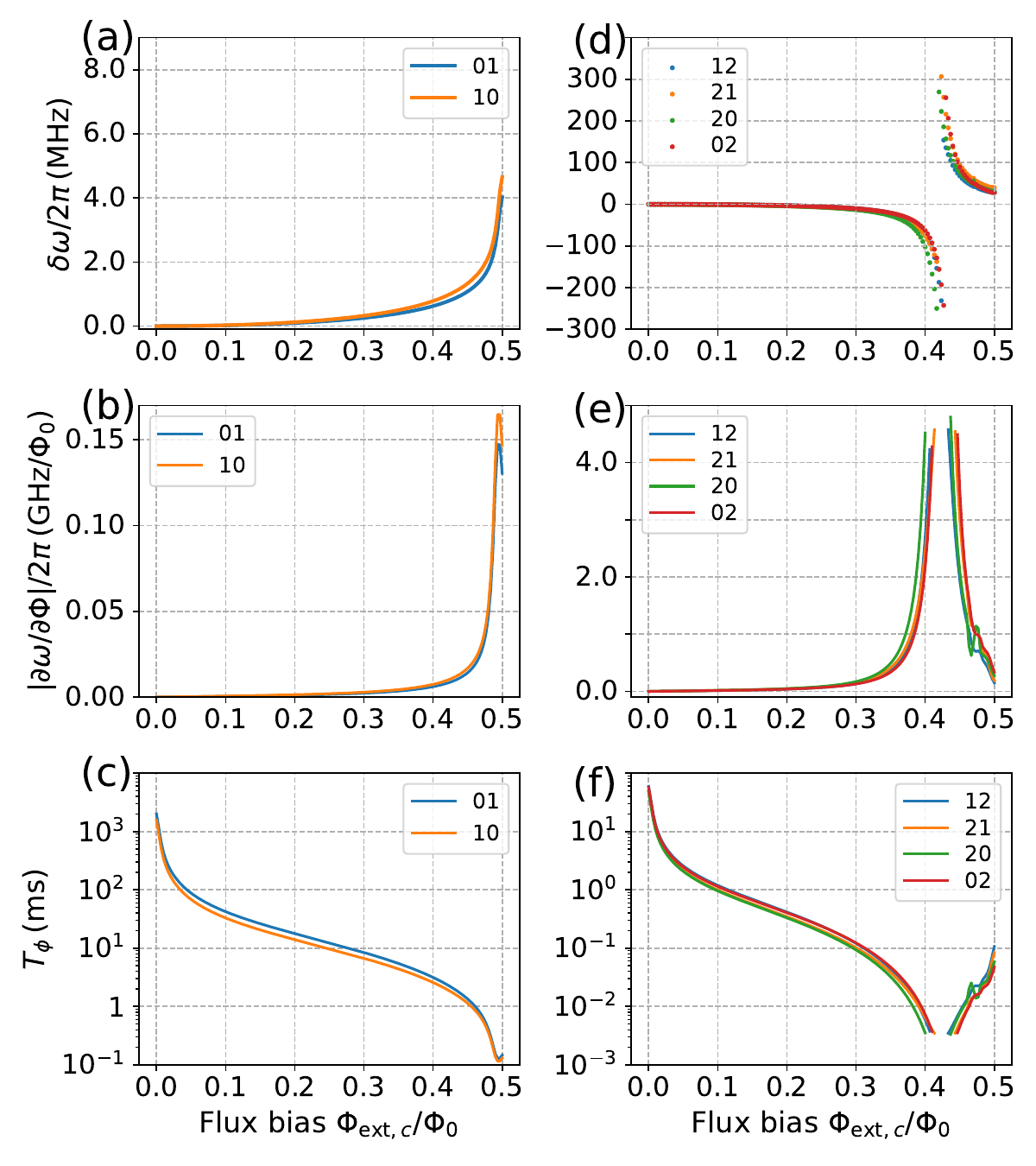}
\end{center}
\caption{Coupler-induced qubit dephasing in the architecture based on the STC. The coupler-induced frequency
shift on the fluxonium levels (a,d), the sensitivity of the energies of these levels to the coupler flux
bias $\partial\omega/\partial\Phi$ (b,e), and the coupler-induced pure dephasing time as functions of the
coupler flux bias (c,f). (a-c) and (d-f) are for the computational and non-computational levels, respectively.
Here, the dephasing times are estimated by assuming 1/f noise with the noise amplitude of $10\,\mu\Phi_{0}$.
We note that sudden jumps or discontinuities in curves
are caused by state labeling failure near avoided crossings.}
\label{STC_decoherence}
\end{figure}

Following the same analytical framework applied to the DTC setup in Sec.~\ref{IF}, we now turn to examine
the coupler-induced decoherence for the architecture based on the STC. Given the general applicability of the
preceding analysis to the present STC case, we focus primarily on summarizing key findings to avoid repetition
while highlighting any setup-specific considerations.

As shown in Fig.~\ref{STC_qubit_overlap} and Figs.~\ref{STC_decoherence}(a-c), the qubit subspace is effectively decoupled from
the STC coupler due to the weak dipole moment of the qubit transition and the low transition frequency, thus heavily suppressing
the Purcell decay and the coupler-induced dephasing. Consequently, similar to the DTC architecture, the STC-based architecture
can also preserve the demonstrated state-of-the-art high coherence in the computational subspace.

However, Figures~\ref{STC_gate_overlap} and~\ref{STC_decoherence}(d-f) reveal that the Purcell decay and coupler-induced dephasing for
non-computational gate states require careful evaluation, particularly when operating the STC near the fluxonium plasmon frequency
to enable strong plasmon interactions and high-speed CZ gates. For the coupler near the plasmon frequency, we observe
approximately $30\%$ state overlap and the corresponding dephasing time of $5\,\rm \mu s$ (assuming $1/f$ noise with the typical noise amplitude of $10\,\mu\Phi_{0}$). For 50-ns CZ gates, this leads to a dephasing error on the order of $10^{-5}$. Importantly, since the reported lifetimes
of these non-computational levels reach $10\,\rm \mu s$~\cite{Ficheux2021,Ding2023}, which is significantly shorter than state-of-the-art
transmon relaxation times ($>100\,\rm \mu s$). Thus, similar to the DTC-based architecture, we conclude that the Purcell decay
does not currently represent the limiting factor for the lifetime of these non-computational levels.

\section{The implementation of CZ gates}\label{III}

Here, we give further detailed descriptions of the microwave-activated CZ gates and
give estimations of the incoherence gate errors from the relaxation and dephasing of the
non-computational gate levels.

\begin{figure}[htbp]
\begin{center}
\includegraphics[keepaspectratio=true,width=\columnwidth]{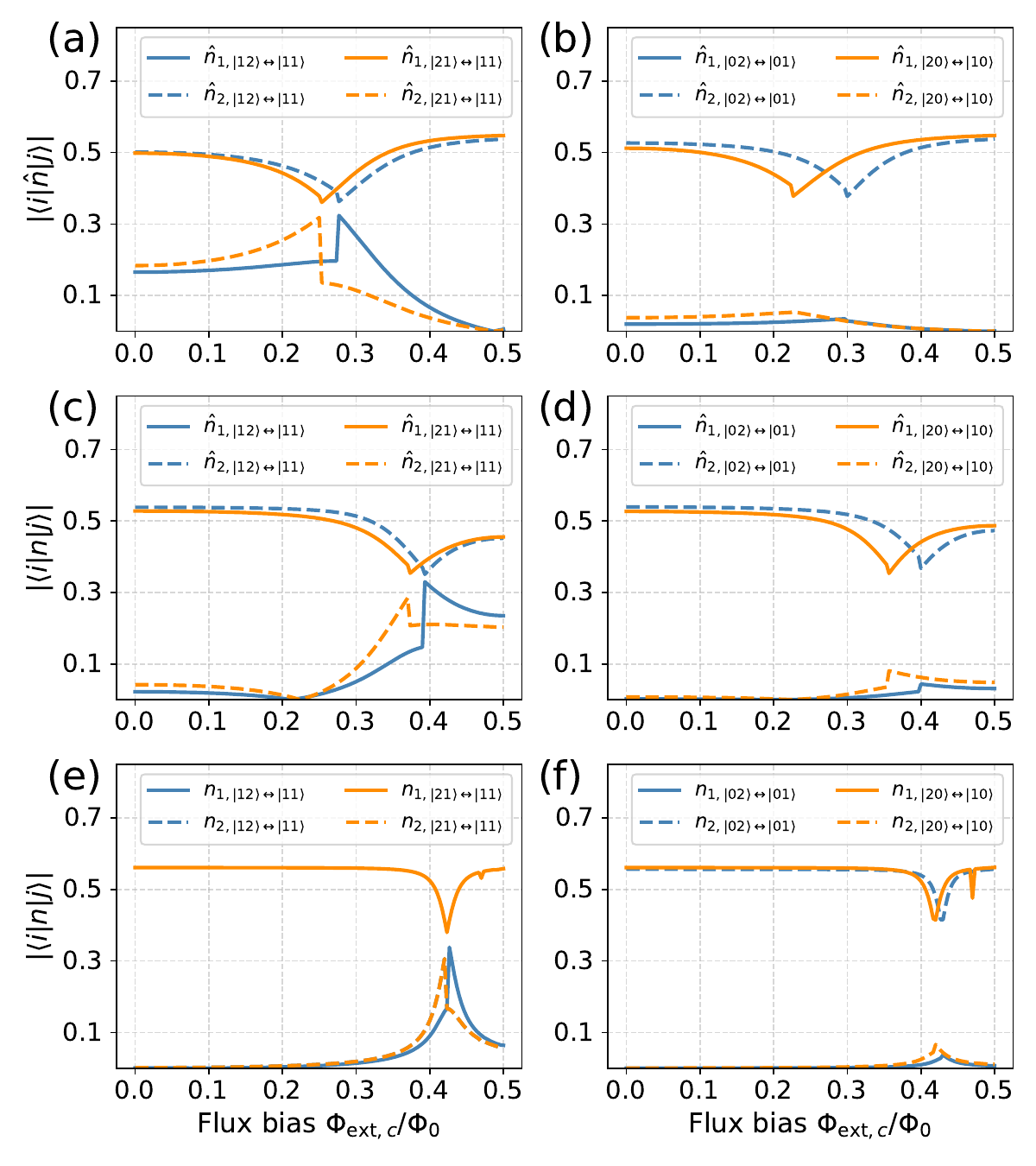}
\end{center}
\caption{Transition magnitudes for the gate transitions ($|12(21)\rangle\leftrightarrow|11\rangle$
and $|02(20)\rangle\leftrightarrow|01(10)\rangle$) versus the coupler flux bias.
(a,b) and (c,d) are for the type-1 and type-2 setup within the DTC-based architecture, respectively, and (e,f)
are for the STC-based architecture. We note that sudden jumps or discontinuities in curves
are caused by state labeling failure near avoided crossings.}
\label{fig19}
\end{figure}

\begin{figure}[htbp]
\begin{center}
\includegraphics[keepaspectratio=true,width=\columnwidth]{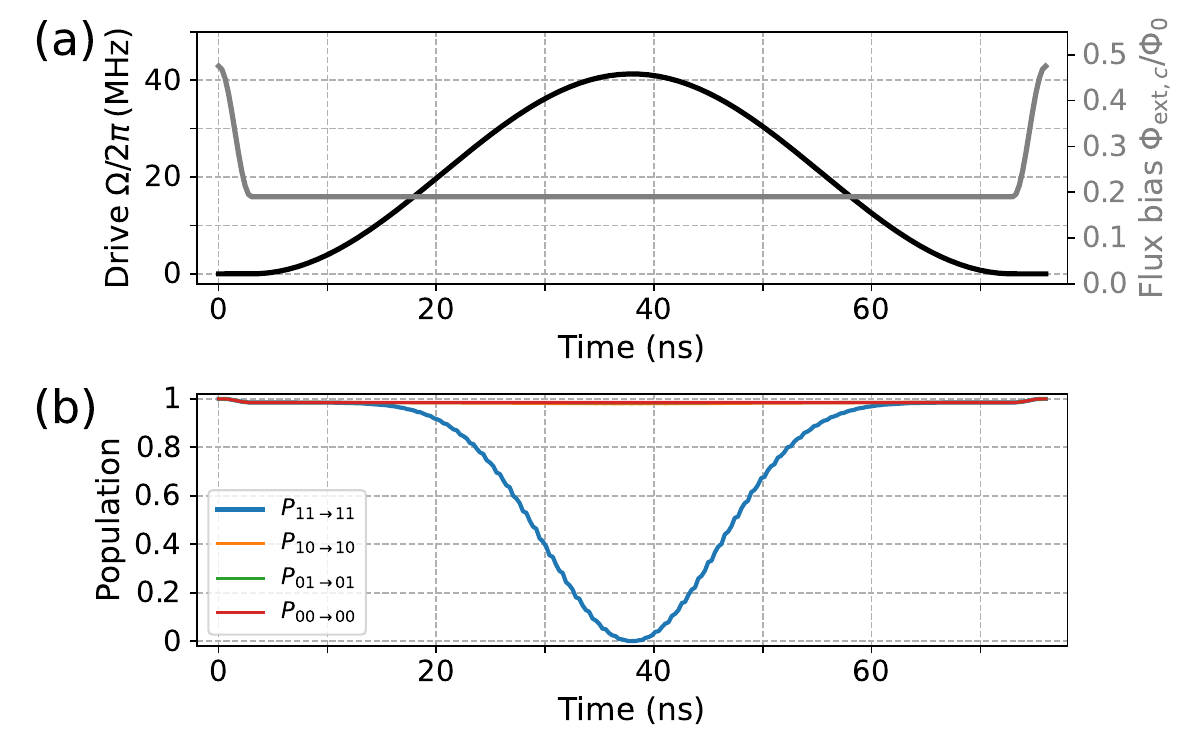}
\end{center}
\caption{(a) A typical control pulse (including the flux biasing pulse and the microwave
drive pulse) for implementing CZ gates within the proposed fluxonium architecture. The
ramp time of flux pulse is 3 ns. (b) The typical system dynamics
during the microwave-activated CZ gate operation. $P_{ij\rightarrow ij}$ denotes the population
in $|ij\rangle$ with the system initialized in the state of $|ij\rangle$.}
\label{fig20}
\end{figure}

\subsection{Microwave-activate CZ gates}\label{IIIA}

As discussed in the main text, in coupled-fluxonium systems, the plasmon interaction can shift
one fluxonium's plasmon frequency shift dependent on the other fluxonium's state. This
conditional shift can be used for selectively driving one particular
transition (gate transition), e.g., $|11\rangle\leftrightarrow|21\rangle$, and the qubit state, e.g., $|11\rangle$, will
accumulate an additional phase of $\pi$ when the system completes a full Rabi oscillation~\cite{Nesterov2018}. In
the proposed architecture, the coupler-mediated plasmon interaction is turned off, i.e., the
coupler is biased at its idle point, for suppressing quantum crosstalk and enabling high-fidelity
single-qubit addressing during the system idle time. Accordingly, the CZ gates can be obtained by firstly
biasing the coupler from its idle point to the interaction point (where the strong plasmon interactions
is activated, e.g., marked by the open circles or squares in Fig.~\ref{fig3} of the main text and in Fig.~\ref{setup_V2}), then
applying a microwave pulse to drive the gate transition and waiting for a full period of the induced Rabi oscillation, and
finally biasing the coupler back to the idle point.

As shown in Fig.~\ref{fig19}, which shows the transition magnitudes versus the coupler
flux bias, the strong state hybridization from the plasmon interaction can enable
the excitation of gate transitions by applying microwave drive to any of the two
fluxoniums. Thus, as in Refs.~\cite{Nesterov2018,Ding2023}, we consider that microwave drives with an identical
amplitude and frequency are applied simultaneously to the two fluxoniums for activating the gate
transitions and the driven Hamiltonian can be modeled by
\begin{equation}
\begin{aligned}\label{eqF1}
H_{d}=\sum_{i=1,2}A(t)\cos(\omega_{d}t+\phi_{i})\hat n_{i},
\end{aligned}
\end{equation}
where $A$, $\omega_{d}$, and $\phi_{i}$ denotes the amplitude, frequency, and the
phase of the drive, respectively. We note that the relative phase between the two drives
are determined by minimizing the period of the activated Rabi oscillation~\cite{Ding2023}.

Here, for illustration purposes, we consider using a flat-top pulse with cosine-shape ramps for
biasing the coupler, i.e.,
\begin{align}
\varphi_{{\rm ext},c}(t)= \varphi_{{\rm ext},c}^{\rm off}+
\begin{cases}
A_{\varphi}\frac{1-\cos{(\pi\frac{t}{t_r}})}{2}  \;, &t\in[0,t_r]\\
A_{\varphi}\;,  &t\in[t_r,t_b-t_r]\\
A_{\varphi}\frac{1-\cos{(\pi \frac{t_b-t}{t_r}})}{2} \;, &t\in[t_b-t_r,t_b]
\end{cases}
\label{eqF2}
\end{align}
where $t_{r}$ and $t_{b}$ denote the pulse ramp time and the full pulse length, and
$\varphi_{{\rm ext},c}^{\rm off}$ denotes the coupler's idle point, and using the cosine-shape
pulse for the microwave drive, i.e.,
\begin{equation}
\begin{aligned}\label{eqF3}
&A(t)=\Omega_{d}\left(1-\cos\frac{2\pi t}{t_{g}}\right).
\end{aligned}
\end{equation}
where $t_{g}$ denotes the gate length (excluding the ramp times for biasing the coupler)
and $\Omega_{d}$ is the peak drive amplitude. We note that due to the near-complete decoupling of
the computational subspace and the coupler, non-adiabatic transitions for qubit states during the flux bias ramping
can almost be neglected. Thus, the non-adiabatic error can in principle
be safely ignored (as confirmed by fact that for CZ gate analysis shown in Fig.~\ref{fig3} of the main text, the
microwave drive parameters are optimized firstly at the coupler interaction
points and then the full pulse including both the flux biasing pulse and the microwave drive pulse are
used for evaluating the gate performance). Here, we select a 3 ns ramp time to accommodate control
electronics constraints in practical systems.

Figure~\ref{fig20}(a) shows the typical control pulse for implementing the CZ gates. To
tune up the CZ gates, the gate parameters, i.e., the drive peak amplitude $\Omega_{d}$ and the
drive frequency $\omega_{d}$ are optimized by minimizing the leakage~\cite{Wood2018} and the
conditional phase error~\cite{Ding2023}. Given the optimized gate parameters, Figure~\ref{fig20}(b) shows
the typical system dynamics during the gate operations. To further evaluate the intrinsic gate
performance (without the consideration of decoherence), the metric of state-average gate
fidelity is used, i.e., up to single-qubit Z phases, the gate fidelity of the
implemented CZ gate is \cite{Pedersen2007}
\begin{equation}
\begin{aligned}\label{eqF4}
F=\frac{{\rm Tr}(\tilde{U}^{\dagger}\tilde{U})+|{\rm Tr}(U_{\rm CZ}^{\dag}\tilde{U})|^{2}}{20},
\end{aligned}
\end{equation}
where $\tilde{U}$ denotes the actual evolution operator, which is truanted to the two-qubit
computational subspace spanned by $\{|00\rangle,|01\rangle,|10\rangle,|11\rangle\}$,
and $U_{\rm cz}$ denote the ideal CZ gate.

\begin{figure}[htbp]
\begin{center}
\includegraphics[keepaspectratio=true,width=\columnwidth]{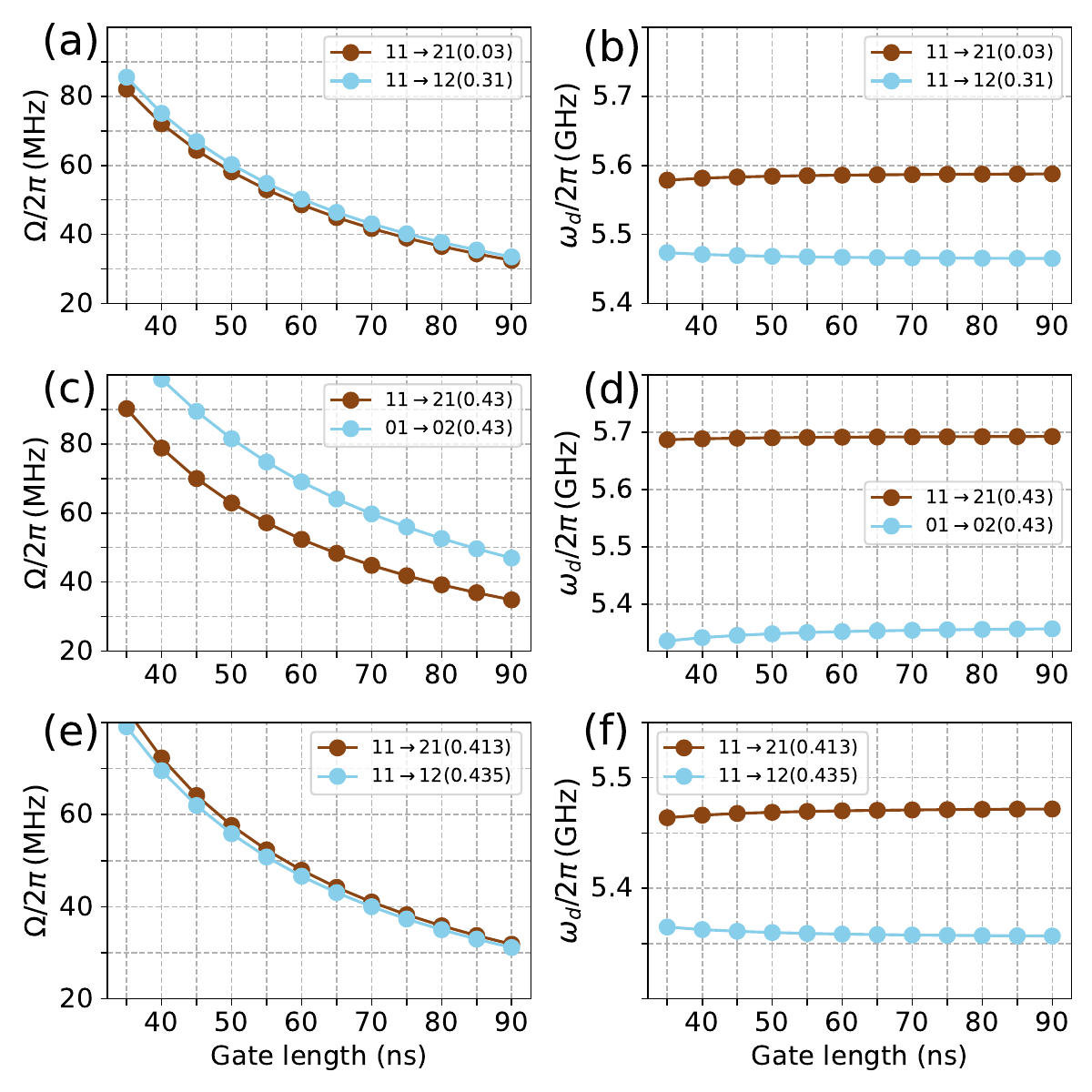}
\end{center}
\caption{The optimized CZ gate parameters (i.e., gate drive amplitudes and drive frequencies) used
for the results shown in Figs.~\ref{fig3}(c) and~\ref{fig3}(g) of the main text and in Fig.~\ref{setup_V2}(e).
(a,b) and (c,d) are for the type-1 and type-2 setup within the DTC-based architecture, respectively, and (e,f)
are for the STC-based architecture.}
\label{fig21}
\end{figure}

Following to the above procedure, we show the gate errors as functions of the gate length
for different gate transitions, see Figs.~\ref{fig3}(c) and~\ref{fig3}(d) of the main text and in Fig.~\ref{setup_V2}(e).
Additionally, Figure~\ref{fig21} shows the the associated gate
parameters, i.e., the peak drive amplitude and the drive frequency.

\begin{figure}[htbp]
\begin{center}
\includegraphics[keepaspectratio=true,width=\columnwidth]{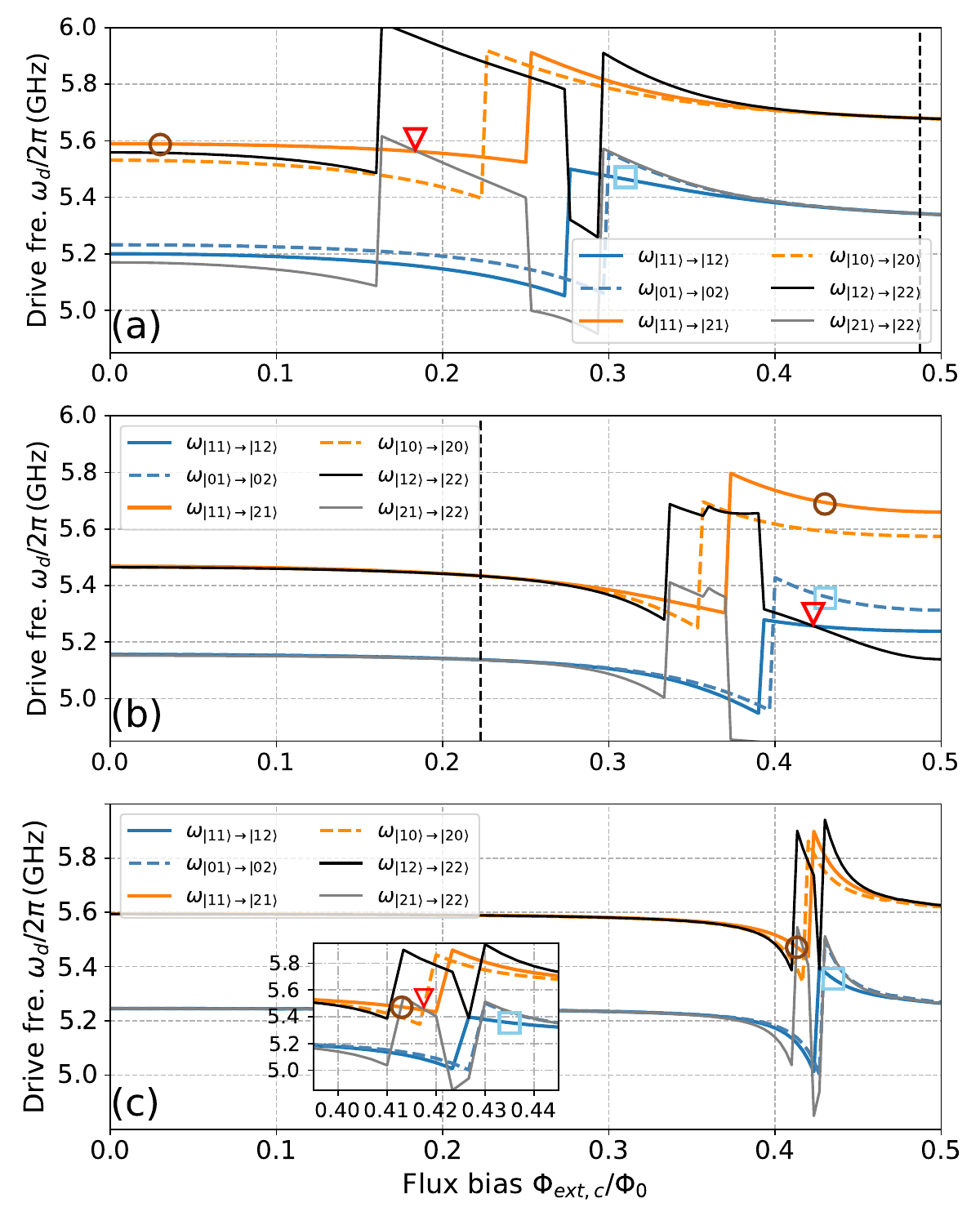}
\end{center}
\caption{Frequencies of the four gate transitions and other leading parasitic transitions
versus the coupler flux bias. (a) and (b) are for type-1 and type-2 setup within the DTC-based
architecture, respectively, and (c) is for the STC-based architecture. Open circles or squares mark the coupler interaction points
used for the CZ gate analysis in the main text and in Fig.~\ref{setup_V2}(e), and red triangle indicate the frequency
collisions from the connection between gate transitions and parasitic
transitions, i.e., $|11\rangle\leftrightarrow|21(12)\rangle\leftrightarrow|22\rangle$. We note that
sudden jumps or discontinuities in curves
are caused by state labeling failure near avoided crossings.}
\label{fig22}
\end{figure}

\subsection{The effect of the $\dot{\Phi}_{{\rm ext},c}$ term}\label{IIIB}

As discussed in Sec.~\ref{IIIA}, implementing CZ gates requires time-dependent flux biasing to turn on plasmon interactions
by shifting the coupler from its idle point to the interaction point and finally back. Our gate analysis is based on the
lumped-element circuit Hamiltonian (Eq.~(\ref{eq1}) of the main text), corresponding to the circuits in Figs.~\ref{fig1}(b) and~\ref{fig1}(c).
Recent studies~\cite{You2019,Riwar2022,Osborne2024,Bryon2023} reveal that quantizing lumped-element circuits
under time-dependent flux generally introduces Hamiltonian terms proportional
to $\dot{\Phi}_{{\rm ext}}$ (involving a gauge transformation), typically neglected in previous works
including ours. While these terms can be eliminated through irrotational gauge choices~\cite{You2019,Osborne2024}, this
approach strictly applies only to lumped-element circuits with well-defined junction capacitances (a well-defined
capacitance is assigned to each Josephson junction). This presents particular challenges for the STC architecture and
the Type-1 setup within DTC-based architecture, where flux biasing through dc SQUID precludes unambiguous capacitance
assignment without detailed geometric considerations~\cite{Riwar2022}. On the contrasty, for the Type-2 setup
within the DTC-based architecture, such capacitance allocation and the irrotational constraint can be
determined, as demonstrated in Ref.~\cite{Campbell2023}.

Thus, here, we turn to analyze the impact of these time-dependent terms on gate dynamics directly, but
qualitatively. As demonstrated in Refs.~\cite{Campbell2023,You2019,Riwar2022,Osborne2024,Bryon2023}, flux biasing
introduces an additional Hamiltonian term proportional to $\dot{\Phi}_{{\rm ext}}$. Specifically, this term takes
the form $\sim \dot{\Phi}_{{\rm ext}} (\hat a_{c}+\hat a_{c}^{\dag})$ for
the STC-based architecture, while for the the Type-1 and Type-2 setup within the DTC-based architecture, the
terms are $\sim \dot{\Phi}_{{\rm ext}} (\hat a_{c1}+\hat a_{c1}^{\dag})(\hat a_{c2}+\hat a_{c2}^{\dag})$
and $\sim \dot{\Phi}_{{\rm ext}} (\hat a_{ci}+\hat a_{ci}^{\dag})$, respectively. Generally, these
terms correspond to coupler drive terms and parametric-activated intermode coupling terms.
For the flux biasing profile in Eq.~(\ref{eqF2}) with the ramp time of 3 ns, the resulting drive and modulation
frequencies ($\sim 167\,\rm MHz$) are significantly detuned from typical coupler modes ($\sim 5\,\rm GHz$).
Given that couplers are generally assumed to be in their ground states before the ramping, these terms will unlikely excite
the coupler modes but only contributing to drive-induced shifts in coupler frequencies during ramping. Moreover, due to the near-complete
decoupling of the computational subspace and the coupler, we expect that these terms does not obviously affect the previous
studied gate dynamics. Accordingly, sudden flux ramping could theoretically excite couplers, e.g., Ref.~\cite{Bryon2023} shows that sudden
flux ramping can excite fluxonium qubits, which is biased near the sweet-spot point with extremely low transition frequency. However,
considering practical control electronics constraints, such sudden ramping is unlikely to be readily achieved
for $\sim 5\,\rm GHz$ coupler modes.

\subsection{Parasitic transitions and frequency collisions}\label{IIIC}

Here, we note that while in principle, any of the four gate
transitions, i.e., $|12(21)\rangle\leftrightarrow|11\rangle$
and $|02(20)\rangle\leftrightarrow|01(10)\rangle$, can be activated for implementing
CZ gates, the frequency collision associated with other parasitic transitions may significantly degrade the
gate performance when the gate transition is near on-resonance with the other parasitic
transitions including both the high-level transitions of the fluxonium and the coupler mode.

Figure~\ref{fig22} shows the gate transitions and other
high-level transitions versus the coupler flux bias. We find that due to strong level repulsion
form the strong plasmon-coupler interactions, frequency collision issues can exist, e.g., from
the pair of the gate transition $|11\rangle\rightarrow |21\rangle$ and the high-level transitions
$|21\rangle\rightarrow |22\rangle$, or the pair of $|11\rangle\rightarrow |12\rangle$ and $|12\rangle\rightarrow |22\rangle$,
as indicated by the red triangles in Figs.~\ref{fig22}(a) and~\ref{fig22}(b). Thinks to the flexibility
in the coupling architecture, such frequency collision issues can be mitigated through either optimized flux
biases or alternative gate transitions, see Fig.~\ref{fig22}.

Furthermore, parasitic transitions may arise from accidental (unexpected) frequency collisions with coupler modes, particularly
when the coupler mode approaches resonance with fluxonium plasmon frequencies. In the STC-based
architecture (see Fig.~\ref{fig3}(g) of the main text), the CZ gate error using the gate transition of $|11\rangle\rightarrow |21\rangle$ exhibits a
decreasing trend with increasing gate length, accompanied by pronounced performance oscillations. These oscillations
originate from the off-resonant transition of $|10\rangle\rightarrow |20\rangle$, generating oscillating leakage errors with the characteristic
period of $1/\sqrt{\delta^2+\Omega^{2}}$ and the amplitude of $\Omega^{2}/(\delta^2+\Omega^{2})$, where $\Omega$ is
the gate drive amplitude and $\delta$ represents the detuning between $|11\rangle\rightarrow |21\rangle$
and $|10\rightarrow 20\rangle$ transitions~\cite{Malekakhlagh2022}. In contrast, the CZ gate error for $|11\rangle\rightarrow |12\rangle$ transition
shows a monotonically decreasing trend without obvious oscillations, as gate error is dominated
by the higher-order (two-photon) $|10,0\rangle\rightarrow |20,1\rangle$ transition that produce slower, weaker
oscillation of leakage. Analogous behavior also occurs in the DTC implementation (see Fig.~\ref{setup_V2}(e)),
where $|11\rangle\rightarrow |21\rangle$ gate performance is limited by the two-photon transition
of $|01,00\rangle\rightarrow |02,01\rangle$. These findings highlight the necessity of careful frequency allocation
for both the coupler modes and the fluxoniums in these implementations.

\begin{figure}[htbp]
\begin{center}
\includegraphics[keepaspectratio=true,width=\columnwidth]{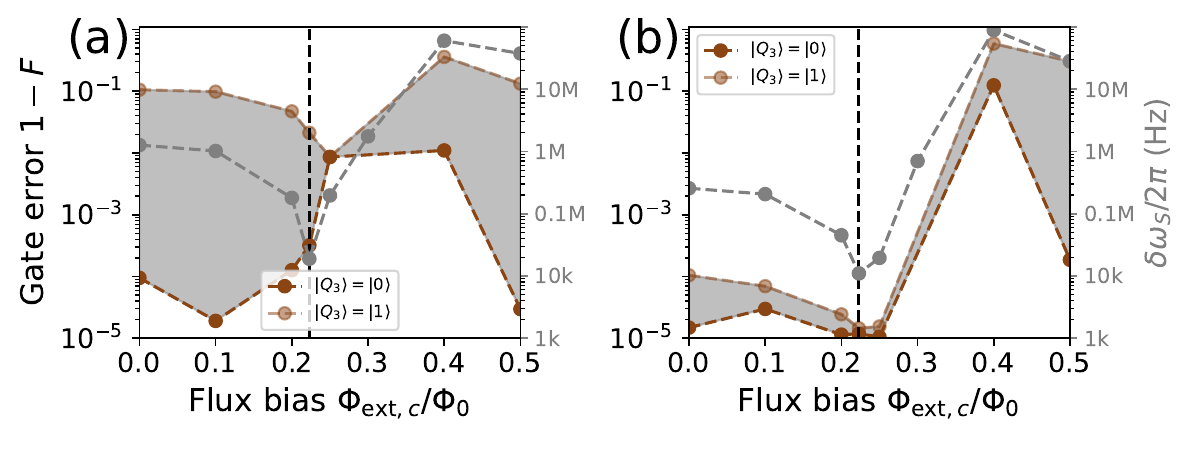}
\end{center}
\caption{Spectator-induced gate errors with varying residuals for the Type-2 setup with the DTC-based
architecture. The grey dots show the shift in the gate frequency due to the
coupling between $Q_{1}$ and the spectator $Q_{3}$ and the brown dots show the gate error with the spectator $Q_{3}$
prepared in different states. (a) The used parameters are the same as in Fig.~\ref{setup_V2}. (b) The used parameters
are the same as in Fig.~\ref{fig17}, i.e., the Set-A of Table~\ref{tab:parameters2}. }
\label{sepctator_v2}
\end{figure}

\subsection{Spectator-induced gate errors}\label{IIID}

To evaluate the scalability of DTC- and STC-based fluxonium architectures, we investigate the influence of residual
spectator-qubit couplings on gate performance (see Figs.~\ref{fig3}(d) and~\ref{fig3}(h) of the main text). Here, we
extend the analysis to the Type-2 setup within the DTC-based architecture. Our extended analysis reveals that even at
the interaction nulling point, substantial gate fidelity degradation can occur through accidental frequency collisions
induced by strong qubit-coupler coupling with small detuning. Thus, analogous to the suppression of residual
couplings, see Sec.~\ref{IE2}, achieving high-performance gates requires maintaining the coupled fluxonium-coupler
system in the dispersive regime to effectively mitigate spectator-induced frequency collisions at nulling points.

Figure~\ref{sepctator_v2}(a) presents the gate error of the CZ gate applied to $Q_{1}$ and $Q_{2}$
while accounting for the interaction between $Q_{1}$ and the spectator $Q_{3}$ (see Fig.~\ref{fig1}(a) of the main text), where the
vertical dashed line marks the flux bias that minimizes their interaction. Notably, even at this nulling point, the
spectator-induced gate error remains substantial. Detailed analysis of the gate dynamics reveals that the error is
dominated by a three-photon transition process involving excitation of the coupler connecting $Q_{1}$
and $Q_{3}$. This unexpected collision originates from the spectral crowding near the fluxonium plasmon transition
when a DTC mode approaches this fluxonium's plasmon transition frequency (see Fig.~\ref{setup_V2}(a)), indicating
operation in a strongly non-dispersive regime. However, as demonstrated in Fig.~\ref{sepctator_v2}(b), detuning the
coupler mode from the plasmon transition (see Fig.~\ref{fig17}) effectively eliminates spectator-induced errors
at the nulling point. These results underscore the importance of coupler frequency engineering
to avoid detrimental frequency collisions involving spectator qubits and coupler modes.

\subsection{Incoherence gate errors from the relaxation and dephasing of non-computational gate levels}\label{IIIE}

Besides the control error including amplitude error and phase error~\cite{Nesterov2018,Ficheux2021,Ding2023}, the
leading gate error could be incoherence errors. We note that as the Hilbert space size of the full system (including two
fluxoniums and couplers) make a full simulation of all decoherence channel less efficient, the following
analysis focuses only on the leading channel. For practical high-coherence fluxonium systems, the dominated incoherence
error should be from the relaxation and dephasing of the non-computational gate states, i.e., $|02(20)\rangle$
and $|12(21)\rangle$~\cite{Abad2023,Didier2019}.

To evaluate gate errors from the relaxation and dephasing of non-computational gate levels, without
loss of generality, here we take the gate transition $|11\rangle\leftrightarrow|21\rangle$ as an illustration.
Accordingly, the evolution operator for the CZ gates can be approximated by
\begin{equation}
\begin{aligned}\label{eqF5}
\hat{U}_{\rm gate}(t)=&|00\rangle\langle 00|+|01\rangle\langle 01|+|10\rangle\langle 10|\\
&+\cos(\Omega t)(|11\rangle\langle 11|+|21\rangle\langle 21|)\\
&-i\sin(\Omega t)(|11\rangle\langle 21|+|21\rangle\langle 11|),
\end{aligned}
\end{equation}
with $\Omega t_{g}=\pi$. Within the framework of the Lindblad master equation, the relaxation and dephasing
of the gate transition can be described by following two collapse operators
\begin{equation}
\begin{aligned}\label{eqF6}
\hat{L}_{21\rightarrow11}=\sqrt{\frac{1}{T_{1}^{21}}}|11\rangle\langle 21|,\quad \hat{L}_{21,\phi}=\sqrt{\frac{2}{T_{\phi,{\rm white}}^{21}}}|21\rangle\langle 21|,
\end{aligned}
\end{equation}
where $T_{1}^{21}$ and $T_{\phi,{\rm white}}^{21}$ denote the relaxation time and the dephasing (white noise) time of the
non-computational level $|21\rangle$, respectively.

Following the procedure given in Ref.~\cite{Abad2023}, the incoherence error for the collapse operator of $\hat{L}$ can
be approximated by
\begin{equation}
\begin{aligned}\label{eqF7}
\epsilon_{\hat{L}} =& \int_{0}^{t_g}dt\left(\frac{{\rm Tr}[\hat{L}^\dag(t)\hat{L}(t)]}{5}-\frac{{\rm Tr}[ \hat{L}^\dag(t)] {\rm Tr}[ \hat{L}(t)]}{20}
\right)
\end{aligned}
\end{equation}
where $\hat{L}(t)=\hat{U}_{\rm gate}^{\dag}(t)\hat{L}\hat{U}_{\rm gate}(t)$ and the trace is only over the states
in the computational subspace. Accordingly, the incoherence gate errors from the relaxation and dephasing are
\begin{equation}
\begin{aligned}\label{eqF8}
&\epsilon_{\hat{L}_{21\rightarrow11}} =\int_{0}^{t_g}\frac{dt}{T_{1}^{21}}\left[\frac{\sin^{2}(\Omega t)}{5} -\frac{\sin^{2}(\Omega t)\cos^{2}(\Omega t)}{20} \right],\\
&\epsilon_{\hat{L}_{21,\phi}} =\int_{0}^{t_g}\frac{2dt}{T_{\phi,{\rm white}}^{21}}\left[\frac{\sin^{2}(\Omega t)}{5} -\frac{\sin^{4}(\Omega t)}{20} \right].
\end{aligned}
\end{equation}

Combining the above two contributions with the dephasing error from the $1/f$ noise~\cite{Didier2019}, the gate fidelity is expressed as
\begin{equation}
\begin{aligned}\label{eqF9}
F=1-\frac{3}{32}\frac{t_{g}}{T_{1}^{21}}-\frac{13}{80}\frac{t_{g}}{T_{\phi,{\rm white}}^{21}}-\frac{13}{80}\left(\frac{t_{g}}{T_{\phi,{\rm 1/f}}^{21}}\right)^{2},
\end{aligned}
\end{equation}
where $T_{\phi,{\rm 1/f}}^{21}$ represent the dephasing time from $1/f$ noise. As mentioned in Sec.~\ref{IF}, the
incoherence gate error from the $1/f$ noise can be less than $10^{-5}$ with $T_{\phi,{\rm 1/f}}^{21}>10\rm\mu s$.
Meanwhile, the incoherence error is below $10^{-3}$ ($10^{-4}$) with the relaxation or dephasing (white noise) time
above $5\,\mu \rm s$ ($50\,\mu\rm s$). We note that further reductions in gate duration and thus error rates may be
achievable through implementation of optimized drive pulses~\cite{Motzoi2009}.


\begin{thebibliography}{99}
\bibitem{note2025} Throughout this study, fluxonium qubits are biased at the half-integer flux quantum unless explicitly stated.
\end{thebibliography}
\end{document}